\documentclass[prd,floats,floatfix,nofootinbib,twocolumn]{revtex4}
\usepackage[dvips]{graphicx}
\usepackage{graphics}
\usepackage{dcolumn}
\usepackage{amssymb}
\usepackage{bm}
\usepackage{amsmath}
\usepackage{color}
\usepackage{epstopdf}

\def\spose#1{\hbox to 0pt{#1\hss}}

\def\lta{\mathrel{\spose{\lower 3pt\hbox{$\mathchar"218$}}
     \raise 2.0pt\hbox{$\mathchar"13C$}}}
\def\gta{\mathrel{\spose{\lower 3pt\hbox{$\mathchar"218$}}
     \raise 2.0pt\hbox{$\mathchar"13E$}}}
\newcommand{\be}{\begin{equation}}
\newcommand{\en}{\end{equation}}
\newcommand{\bea}{\begin{eqnarray}}
\newcommand{\ena}{\end{eqnarray}}

\begin{document}

\title{Maxwell Fields in Boosted Kerr Black Holes}

\author{Rafael F. Aranha$^{a}$\footnote{rafael.fernandes.aranha@uerj.br}, Carlos E. Cede\~no M.$^{b}$\footnote{cecedeno@cbpf.br}, Rodrigo Maier$^{a}$\footnote{rodrigo.maier@uerj.br}, Ivano Dami\~ao Soares$^{b}$\footnote{ivano@cbpf.br}
\vspace{0.5cm}}

\affiliation{$^a$Departamento de F\'isica Te\'orica, Instituto de F\'isica, Universidade do Estado do Rio de Janeiro,\\
Rua S\~ao Francisco Xavier 524, Maracan\~a,\\
CEP20550-900, Rio de Janeiro, Brazil\\
\\
$^b$Centro Brasileiro de Pesquisas F\'{\i}sicas -- CBPF, \\ Rua Dr. Xavier Sigaud, 150, Urca,
CEP 22290-180, Rio de Janeiro, Brazil}

%\date{\today}

\begin{abstract}

The spacetime of a boosted Bondi-Sachs rotating black hole is considered as a proper background to examine electromagnetic configurations connected to analytic solutions of Maxwell equations. In our analysis, we first use the Bondi-Sachs transformations in order to bring the boosted rotating black hole metric into the Kerr-Schild form, from which zero angular momentum observers (ZAMOs) are constructed via the ADM formalism. In Kerr-Schild coordinates we obtain the Killing fields as sources of Maxwell electrodynamics, and we fix a ZAMO in order to evaluate the components of the electric and magnetic fields, from which we obtain nonsingular patterns of an eventual momentum-energy emission of a boosted Kerr-Schild black hole. Distinct patterns are examined and discussed in the case of variations of the boost parameter $\gamma$. We extend our analysis by considering the nonsingular electromagnetic emission in the framework of a boosted Bondi-Sachs rotating black hole, as it moves at relativistic speeds. We also discuss possible mechanisms that may resemble magnetospheres of rotating boosted black holes and give rise to hydromagnetic flows from accretion discs and to the production of jets.
\end{abstract}
\maketitle
\section{Introduction}
The recent direct observations of the gravitational wave emission from binary black hole mergers by the LIGO Scientific Collaboration and the Virgo Collaboration \cite{1}-\cite{4} established that the initial black holes of each binary had mass ratios $\alpha$'s ranging from $\alpha \simeq 0.8$ to $\simeq 0.53$. In this sense the remnant black hole description must contain additional parameters -- boost parameters -- connected to its motion relative to the observation frame\cite{TT}.
% of the remnant 

%
\par The boosted black hole solution can actually be a natural set for astrophysical processes related to the asymmetry of the ergosphere and to the electrodynamics effects that result from the rotating black hole moving at relativistic speeds in a direction coinciding or not with the rotation axis. This is the case of an astrophysical configuration in which external gas can support electric currents that create large-scale magnetic fields. Motion of black holes in this externally supplied magnetic field can then lead to an electromagnetic extraction of energy.
\par Qualitatively, there are two distinct possibilities for the electromagnetic extraction of energy from spiralling black holes.
In the first, the system of two orbiting black holes possesses nonzero angular momentum, which induces rotation of spacetime. Rotating spacetimes can generate electromagnetic outflows, in a manner similar to the classical Faraday disk. This is the physics
behind the Blandford-Znajek process\cite{Blandford} of extracting the rotational power of a black hole. This mechanism is also known as the Faraday disk mechanism. It has also been considered by Lyutikov\cite{Lyutikov}, where the total energy loss from a system of merging black holes is a sum of two components, one due to the rotation of spacetime driven by the nonzero angular momentum in the system, and the other due to the linear motion of the black holes through the magnetic field.
In the second, Morozova et al. \cite{morozova} derived analytic solutions of the Maxwell equations for a rotating black hole moving at constant speed in an asymptotically uniform magnetic test field; electromagnetic energy losses computed from charged particles accelerated along the magnetic field lines constitute a numerical estimate that approximates numerical relativity calculations in a force-free magnetosphere.
\par Apart from the above mentioned processes, it has also been shown\cite{penna0} that black holes can also power jets as long as they 
carry linear momentum and/or orbital momentum. Such jets are proportional to the black hole's velocity, mass and magnetic flux.
It is argued that such jets extract kinetic energy stored in the black hole's motion and that this mechanism
is analogous to that of the energy extraction
from Kerr black holes.
\par In the vein of the above calculations we intend here to examine how efficiently the rotational energy of an isolated boosted Kerr black hole can be extracted by a magnetic field even in the absence of currents and charges. The boost can annulate the Meissner effect\cite{meissner}, allowing instead for jets connected to the mechanism of Blandford and Znajek \cite{Blandford}.
\par
The spacetime of the boosted Kerr black hole is here described as a solution of Einstein's equations at the future null infinity, in the Bondi-Sachs (B-S) coordinates \cite{ids1}, as well as in Robinson-Trautman (R-T) and Kerr-Schild (K-S) coordinates \cite{TT,ids3}.
We organize the paper as follows. In the next section we bring the boosted rotating black hole metric \cite{ids1} into the Kerr-Schild form in order to fix a proper observer (ZAMO). In section III we build a Maxwell test field based on spacetime isometries using Killing vectors. All the components of the electromagnetic field -- with respect to the ZAMO -- are evaluated. Section IV is devoted to numerical simulations of magnetic fields in boosted rotating black holes in the K-S frame. Analogous simulations for the case of a B-S frame are carried out in section V. Final remarks and discussions on future developments are the object of section VI.

\section{The Boosted Rotating Black Hole Metric in the Kerr-Schild Form}
Let us consider the metric of a boosted rotating black hole in which the boost is along the axis of rotation with respect to an asymptotic Lorentz frame at future null infinity. In its original form \cite{ids1} the line element in Bondi-Sachs coordinates $(U, R, \Theta, \Phi)$ is expressed as
\begin{eqnarray}
\label{e1}
\nonumber
&&ds^2=(R^2+\Sigma^2(\Theta))(d\Theta^2+\sin^2{\Theta}d\Phi^2)\\
\nonumber
&-&2(dU+\omega_b(\Theta) \sin^2{\Theta}d\Phi)\Big(dR-\frac{\omega_b(\Theta) \sin^2{\Theta}}{K^2(\Theta)}d\Phi\Big) \\
\nonumber
&-& (dU+\omega_b(\Theta) \sin^2{\Theta}d\Phi)^2
\Big(\frac{1}{K^2(\Theta)}-\frac{2m_b(\Theta) R}{R^2+\Sigma^2(\Theta)}\Big)\\
&+&\mathcal{O}\Big(\frac{1}{R^2}\Big),
\end{eqnarray}
where
\begin{eqnarray}
\label{e2}
K(\Theta)&=&\cosh{\gamma}+\sinh{\gamma}\cos{\Theta},\\
\Sigma(\Theta)&=&\frac{\omega_b(\Theta)}{K(\Theta)}(\sinh{\gamma}+\cosh{\gamma}\cos{\Theta}),\\
\omega_b (\Theta)&=&\frac{w}{K(\Theta)}, ~~m_b(\Theta)=\frac{m_0}{K^3(\Theta)}.
\end{eqnarray}
Here $K(\Theta)$ defines the Lorentz boost of the Bondi-Sachs group \cite{BS1,BS2} and $\gamma$ is the boost parameter. Also $\omega$ is the rotation parameter and $m_0$ is the black hole mass. From (\ref{e1}) it is easy to see that the leading components of the Einstein tensor are of the order $\mathcal{O}(R^{-2})$ or higher so that at future null infinity $G_{\mu\nu}\simeq 0$.
\par
Applying the Bondi-Sachs transformations \cite{aranha,kramer}
\begin{eqnarray}
\label{e22}
\nonumber
d U/d u \sim K(\Theta),\\
R \sim r/K(\Theta), \\
\nonumber
d R/d r  \sim 1/K(\Theta),
\end{eqnarray}
so that  asymptotic Bondi-Sachs conditions are satisfied, the metric (\ref{e1}) can be rewritten in terms of the Robinson-Trautman coordinates $(u, r, \Theta,\Phi)$ as
%
%\begin{widetext}
\begin{eqnarray}
\label{mrt}
\nonumber
ds^2=-\Big[1-\frac{2r m_b(\Theta)  K^3(\Theta)}{r^2+\tilde{\Sigma}^2(\Theta)}\Big]du^2-2du dr~~~~~~~~~~\\
\nonumber
+\frac{4rm_b(\Theta)K^2(\Theta)\omega_b(\Theta) \sin^2{\Theta}}{r^2+\tilde{\Sigma}^2(\Theta)}dud\Phi~~~~~~~~~~\\
-\frac{2\omega_b(\Theta)\sin^2\Theta}{K(\Theta)}drd\Phi +\frac{r^2+\tilde{\Sigma}^2(\Theta)}{K^2(\Theta)}d\Theta^2\\
\nonumber
+\sin^2\Theta\Big[\frac{r^2+\tilde{\Sigma}^2(\Theta)+\sin^2\Theta\omega_b(\Theta)}{K^2{(\Theta)}}~~~~~~~~~~~~~~~~~~\\
\nonumber
+\frac{2rK(\Theta)m_b(\Theta)\omega_b(\Theta)\sin^2\Theta}{r^2+\tilde{\Sigma}^2(\Theta)}\Big]d\Phi^2\\
\nonumber
+ \mathcal{O}\Big(\frac{1}{r^2}\Big),
\end{eqnarray}
%\end{widetext}
%
where $\tilde{\Sigma}(\Theta)\equiv K(\Theta)\Sigma(\Theta)$.
\begin{figure*}
	\begin{center}		
		%\hspace{0.5cm}
		%\vspace{0.0cm}
		%\begin{tabular}{ll}
		\includegraphics[width=5.0cm]{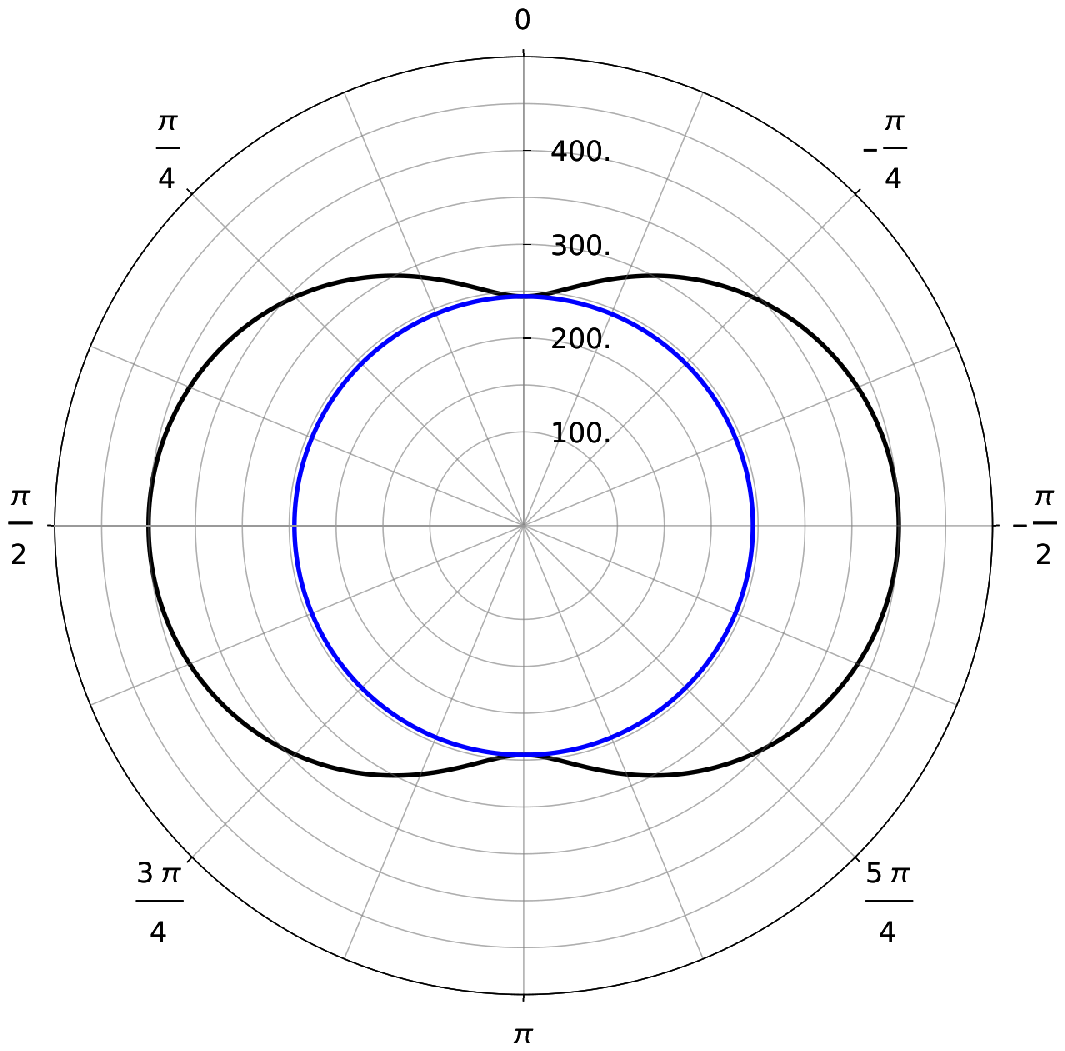}\qquad %&
		\includegraphics[width=5.0cm]{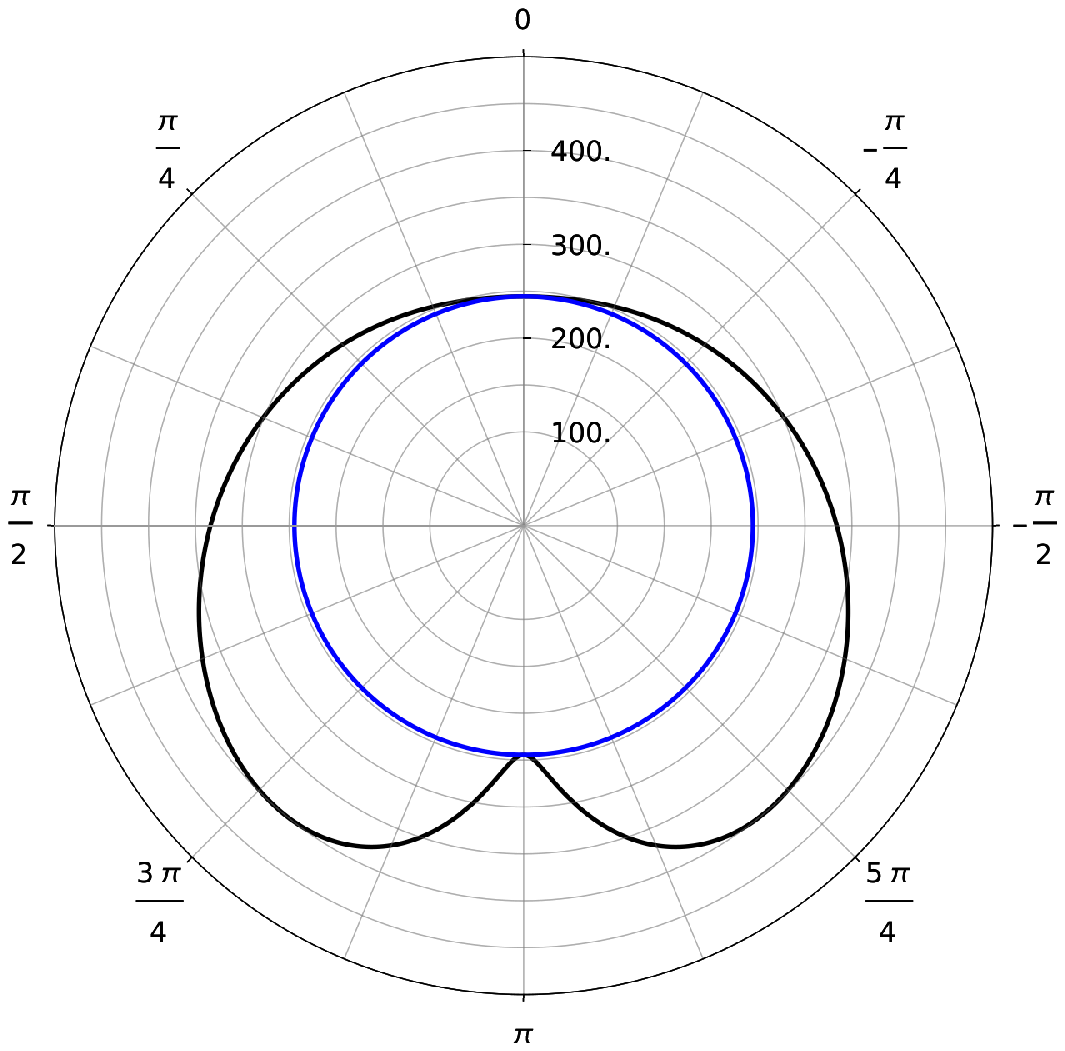}
	   %\end{tabular}       		 
		\caption{%\small{
				Plots of the sections of the ergospheres (black) and the horizons (blue) of a K-S black hole (\ref{ks}),
				by a plane containing the $z$-axis, which is the rotation axis of the black holes. The parameters used in the plots
				are $m_0=200$, $w_0=195$ with  $\gamma=0$ (left) and $\gamma=1.0$ (right). We note the deformation of
				the ergosphere for a nonzero boost parameter $\gamma$.%}
		}
		\label{ergoKS}
	\end{center}
\end{figure*}
%%%%%%%%%%%%%%%%%%%%%%%%%%%%%%%%%%%%%%%%%%%%%%%%%%%%%%%%%%%%%%%%%%%%%%%%%%%%%%%%%%%%%%%%%%%%%%%%%%%%%%%%%%%
%
Finally, in order to fix a preferred $(1+3)$ foliation of the spacetime via the ADM formalism, we take into account the transformation $u=t-r$. In this case the geometry (\ref{mrt}) can be rewritten in Kerr-Schild coordinates $(t, r, \Theta,\Phi)$ as
%
%\begin{widetext}
\begin{eqnarray}
\label{ks}
\nonumber
ds^2=-\Big[1-\frac{2r m_b(\Theta)  K^3(\Theta)}{r^2+\tilde{\Sigma}^2(\Theta)}\Big]dt^2~~~~~~~~~~~~~~~~~~~~~~~~~~\\
\nonumber
-\frac{4rK^3(\Theta)m_b(\Theta)}{r^2+\tilde{\Sigma}^2(\Theta)}dtdr~~~~~~~~~~~~~~~~~~~~~~~~~~~\\
\nonumber
+\frac{4rm_b(\Theta)K^2(\Theta)\omega_b(\Theta) \sin^2{\Theta}}{r^2+\tilde{\Sigma}^2(\Theta)}dtd\Phi~~~~~~~\\
\nonumber
+\Big[1+\frac{2r m_b(\Theta)  K^3(\Theta)}{r^2+\tilde{\Sigma}^2(\Theta)}\Big]dr^2~~~~~~~~~~~~~\\
-2\omega_b(\Theta)\sin^2\Theta\Big[\frac{1}{K(\Theta)}+\frac{2rK^2(\Theta)m_b(\Theta)}{r^2+\tilde{\Sigma}^2(\Theta)}\Big]drd\Phi~~~\\
\nonumber
+\frac{r^2+\tilde{\Sigma}^2(\Theta)}{K^2(\Theta)}d\Theta^2~~~~~~~~~~~~~~~~~~\\
\nonumber
+\sin^2\Theta\Big[\frac{r^2+\tilde{\Sigma}^2(\Theta)+\sin^2\Theta\omega_b(\Theta)}{K^2{(\Theta)}}~~~~~~~~~~~~~~~~~~~~~~\\
\nonumber
+\frac{2rK(\Theta)m_b(\Theta)\omega_b(\Theta)\sin^2\Theta}{r^2+\tilde{\Sigma}^2(\Theta)} \Big]d\Phi^2+ \mathcal{O}\Big(\frac{1}{r^2}\Big).
\end{eqnarray}
%\end{widetext}
%
The leading components of the Einstein tensor of (\ref{ks}) are of the order $\mathcal{O}(r^{-2})$ or higher so that at future infinity $G_{\mu\nu}\simeq 0$, as should be expected.
\par In Figs. \ref{ergoKS} we plot respectively the sections of the ergosphere $R_{stat}$ (black line) and of the event horizon $R_{H}$ (blue line) by a plane containing the $z$-axis. In the second case the ergosphere is deformed due to a nonvanishing Lorentz boost present in (\ref{ks}).

In order to fix a zero angular momentum observer (ZAMO) we evaluate the lapse function $N$, the shift $N^i$ and the spatial metric $\gamma_{ij}$ of (\ref{ks}). Writing the ADM decomposition of (\ref{ks}) as
\begin{eqnarray}
\label{adm}
ds^2=-N^2 dt^2+\gamma_{ij}(dx^i-N^idt)(dx^j-N^jdt),
\end{eqnarray}
the lapse function, the shift and the spatial metric read
\begin{widetext}
\begin{eqnarray}
\label{lssm}
N&=&\pm\sqrt{1+2rK^3(\Theta)m_b(\Theta) \Big[\frac{1}{r^2+2rK^3(\Theta)m_b(\Theta)+\tilde{\Sigma}^2(\Theta)}-\frac{2}{r^2+\tilde{\Sigma}^2(\Theta)}   \Big]}~,\\
N^i&=&\frac{2rK^3(\Theta)m_b(\Theta)}{r^2+2rK^3(\Theta)m_b(\Theta)+\tilde{\Sigma}^2(\Theta)}\delta^i_{~r}~,\\
\nonumber
\gamma_{ij}&=&1+\frac{2rK^3(\Theta)m_b(\Theta)}{r^2+\tilde{\Sigma}^2(\Theta)}\delta^r_{~i}\delta^r_{~j}
-\omega_b(\Theta)\sin^2{\Theta}\Big[\frac{1}{K(\Theta)}+\frac{2rK^2(\Theta)m_b(\Theta)}{r^2+\tilde{\Sigma}^2(\Theta)}\Big]\delta^r_{~i}\delta^\Phi_{~j}
+\frac{r^2+\tilde{\Sigma}^2(\Theta)}{K^2(\Theta)}\delta^\Theta_{~i}\delta^\Theta_{~j}\\
&+&\sin^2(\Theta)\Big[\frac{r^2+\omega_b^2(\Theta)\sin^2{\Theta}+\tilde{\Sigma}^2(\Theta)}{K^2(\Theta)}
+\frac{2rK(\Theta)m_b(\Theta)\omega^2_b(\Theta)\sin^2{\Theta}}{r^2+\tilde{\Sigma}^2(\Theta)}\Big]\delta^\Phi_{~i}\delta^\Phi_{~j}~.
\end{eqnarray}
\end{widetext}

For the case of a radially falling ZAMO\cite{takahashi}, the $4$-velocity $u^\mu$ is given by
\begin{eqnarray}
\label{zamo}
u^\mu=\Big(\frac{1}{N}, \frac{N^r}{N}, 0, 0\Big).
\end{eqnarray}
In the following section we will examine electrodynamic configurations in a boosted black hole as measured by ZAMO observers assuming a Maxwell
test field in the background (\ref{ks}).
However, at the final part of the paper, we will approach the electrodynamics of the boosted BH as described by Bondi-Sachs asymptotic observers at the future null infinity of (\ref{e1}).

\section{Killing Vectors as a Test Maxwell Field}

It is well known that Killing vectors are solutions of Maxwell equations for vacuum spacetimes\cite{papapetrou,wald}. In fact, let $K_{\mu}$ be the vector potential that defines the Faraday tensor $F_{\mu\nu}$, namely,
\begin{eqnarray}
\label{e5}
F_{\mu\nu}=\nabla_{\nu}K_{\mu}-\nabla_{\mu}K_{\nu}.
\end{eqnarray}
Assuming that $K_{\mu}$ is also a Killing vector, we obtain
\begin{eqnarray}
\label{e6}
F_{\mu\nu}=-2\nabla_{\mu}K_{\nu},
\end{eqnarray}
so that Maxwell equations in vacuum spacetime read
\begin{eqnarray}
\label{e7}
\nabla_{\nu}F^{\mu\nu}=\square K^{\mu}=0.
\end{eqnarray}
However,
\begin{eqnarray}
\label{e8}
\nabla_{\gamma}\nabla_{\mu}K_{\nu}-\nabla_{\mu}\nabla_{\gamma}K_{\nu}=-R^{\sigma}_{~\nu\mu\gamma}K_{\sigma},
\end{eqnarray}
where $R^{\sigma}_{~\nu\mu\gamma}$ is the Riemann tensor. Now, cyclic permutations of (\ref{e8}) yield
\begin{eqnarray}
\label{e9}
\nabla_{\mu}\nabla_{\nu}K_{\gamma}-\nabla_{\nu}\nabla_{\mu}K_{\gamma}=-R^{\sigma}_{~\gamma\nu\mu}K_{\sigma},
\end{eqnarray}
and
\begin{equation}
\label{e10}
\nabla_{\nu}\nabla_{\gamma}K_{\mu}-\nabla_{\gamma}\nabla_{\nu}K_{\mu}=-R^{\sigma}_{~\mu\gamma\nu}K_{\sigma}.
\end{equation}
\\
Therefore, after subtracting (\ref{e10}) from the sum of (\ref{e8}) with (\ref{e9}) we obtain
\begin{eqnarray}
\label{e11}
\nonumber
2\nabla_{\mu}\nabla_{\nu}K_{\gamma}&=&-(R^{\sigma}_{~\nu\mu\gamma}+R^{\sigma}_{~\gamma\nu\mu}-R^{\sigma}_{~\mu\gamma\nu})
K_{\sigma}\\
&\equiv& 2 R^{\sigma}_{~\mu\gamma\nu} K_{\sigma}.
\end{eqnarray}
Hence
\begin{eqnarray}
\label{e12}
\square K_\gamma = R^\sigma_{~\gamma} K_\sigma,
\end{eqnarray}
and for a vacuum spacetime equation (\ref{e7}) is automatically satisfied.
\par

The above geometry (\ref{ks}) allows us to identify the two Killing vectors
$\Gamma^\mu=\delta^\mu_{~t}$ and $\Psi^\mu=\delta^{\mu}_{~\Phi}$.
We now evaluate the Faraday tensor for both Killing vectors in Kerr-Schild coordinates. In the present section, primes denote the derivative with respect to $\Theta$.
\par For the Killing vector $\Gamma^\mu$ the nonvanishing components of $F_{\mu\nu}$ are given by (see (\ref{e6}))
\begin{eqnarray}
\label{e13}
{^t}F_{02}&\simeq& - {^t}F_{12} \simeq \frac{2}{r}(K^3 m_b)^\prime~,\\
{^t}F_{23}&\simeq& -\frac{4 m_b\omega_b K \sin\Theta}{r}(\sin{\Theta}K)^\prime.
\end{eqnarray}
On the other hand, for the Killing vector $\Psi^\mu$ the nonvanishing components of $F_{\mu\nu}$ read
%
%\begin{widetext}
\begin{eqnarray}
\label{e15}
{^\Phi}F_{02} &\simeq& 2\frac{m_b\omega_b}{r} (K^2 \sin^{2}\Theta)^{\prime },\\
\nonumber
{^\Phi}F_{12}&\simeq& \frac{2\omega_b \sin^2{\Theta}}{K^2}(K^\prime - K\cot{\Theta})\\
&&~~~~~~~~~~- \frac{4 m_b\omega_b\sin\Theta}{r}(K \sin{\Theta})^{\prime} ,\\
{^\Phi}F_{13}&\simeq&-\frac{2r\sin^2\Theta}{K^2},\\
\nonumber
{^\Phi}F_{23}&\simeq& \frac{2r^2\sin^2{\Theta}}{K^3}(K^\prime- K\cot{\Theta})\\
\nonumber
&-&\frac{2\sin{\Theta}}{K^5}[K^3\omega_b^2\sin\Theta\sin{2\Theta}-2K^2K^\prime \omega_b^2 \sin^3\Theta \\
&&~~~~~~~+ K^5\Sigma~(\Sigma\cos\Theta +\Sigma^\prime\sin\Theta)]\\
\nonumber
&&~~~~~~~~~~+\frac{8m_b\omega^2_b\sin^4\Theta}{r}(K^\prime-K\cot{\Theta}).
\end{eqnarray}
%\end{widetext}
%

By defining the Hodge dual by ${\cal F}^{\mu\nu}=\frac{1}{2}\epsilon^{\mu\nu\alpha\beta}F_{\alpha\beta}$ we obtain their correspondence
\begin{eqnarray}
\label{e14}
{^t}{\cal F}^{01}&\simeq& {^t}F_{23},\\
{^t}{\cal F}^{03}&\simeq& {^t}F_{12},\\
{^t}{\cal F}^{13}&\simeq& -{^t}F_{02},
\end{eqnarray}
and
\begin{eqnarray}
{^\Phi}{\cal F}^{01}&\simeq& {^\Phi}F_{23},\\
{^\Phi}{\cal F}^{02}&\simeq& -{^\Phi}F_{13},\\
{^\Phi}{\cal F}^{03}&\simeq& {^\Phi}F_{12},\\
{^\Phi}{\cal F}^{13}&\simeq& -{^\Phi}F_{02}.
\end{eqnarray}

We are now in a position to evaluate the electric and magnetic fields using their usual definitions, ${\cal E}^\mu=F^\mu_{~~\nu}u^\nu~~ {\rm and}~~ {\cal B}^\mu={\cal F}{^\mu}_{\nu}u^\nu$.
We obtain
\begin{eqnarray}
\label{em1}
{^t}{\cal E}^\mu \simeq {^\Phi}{\cal E}^\mu \simeq 0.
\end{eqnarray}
On the other hand for the magnetic fields the leading components are

%\begin{widetext}
\begin{eqnarray}
\label{mf}
\nonumber
{^t}{\cal B}^r&\simeq& -\frac{2K\sin{\Theta}}{r}\{\omega_b[2m_b(K\sin{\Theta})^\prime~~~~~~~~~~~~~~~~~~\\
&&+ Km_b^\prime\sin{\Theta}]+Km_b\omega_b^\prime\sin{\Theta}\},\\
\label{mf1}
{^t}{\cal B}^\Phi&\simeq& -\frac{2K^2}{r}(3m_b K^\prime+m_b^\prime K), 
\end{eqnarray}
and
\begin{eqnarray}
\nonumber
\label{mf2}
\nonumber
{^\Phi}{\cal B}^r&\simeq& \frac{2r^2\sin^2{\Theta}}{K^3}(K^\prime-\cot{\Theta}K)\\
\nonumber
&+&2r m_b\sin{\Theta}(K\cos{\Theta}-K^\prime\sin{\Theta})\\
\nonumber
&-&\frac{\sin{\Theta}}{K^3}[11m_b^2K^6(K\cos{\Theta}-K^\prime\sin{\Theta})\\
\nonumber
&-&2\omega_b^2K^\prime\sin^3{\Theta}
+2K\omega_b\sin^2{\Theta}
\\
\nonumber
&\times&(2\omega_b\cos{\Theta}+\omega_b^\prime\sin{\Theta})+2K^3\Sigma(\Sigma\sin{\Theta})^\prime]\\
\label{mf3}
&+&\frac{\sin{\Theta}}{r}\{13K^6m_b^3(K^\prime\sin{\Theta}-K\cos{\Theta})\\
\nonumber
&+&4m_b\omega_b^2K^\prime\sin^3{\Theta}-2K^2K^\prime m_b\Sigma^2\sin{\Theta}\\
\nonumber
&+&2K\omega_b\sin^2{\Theta}[\omega_bm_b^\prime\sin{\Theta}\\
\nonumber
&+&m_b(2\cos{\Theta}\omega_b+\omega_b^\prime\sin{\Theta})]\\
\nonumber
&-&K^3m_b\sin{\Theta}(\Sigma^2)^\prime\},\\
\nonumber
{^\Phi}{\cal B}^\Theta &\simeq& \frac{2r\sin{\Theta}}{K^2}-2Km_b\sin^2{\Theta}\\
&+&\frac{11K^4m_b^2\sin^2{\Theta}}{r},
\\
\nonumber
\label{mf4}
{^\Phi}{\cal B}^\Phi &\simeq& \frac{\sin{\Theta}}{K^2}[\omega_b K^\prime \sin{\Theta}-K(2\omega_b\cos{\Theta}+\omega_b^\prime
\sin{\Theta})]\\
\nonumber
&-&\frac{K\sin{\Theta}}{r}\{5m_b\omega_bK^\prime\sin{\Theta}+K[2\omega_b(m_b\sin{\Theta})^\prime\\
&+&m_b\omega_b^\prime\sin{\Theta}]\}.
\end{eqnarray}
%\end{widetext}
%
\begin{figure*}
	\begin{center}
		\begin{tabular}{ccc}
			\begin{minipage}{5cm}
			\vspace{-5.5cm}	
			\includegraphics*[height=4.5cm]{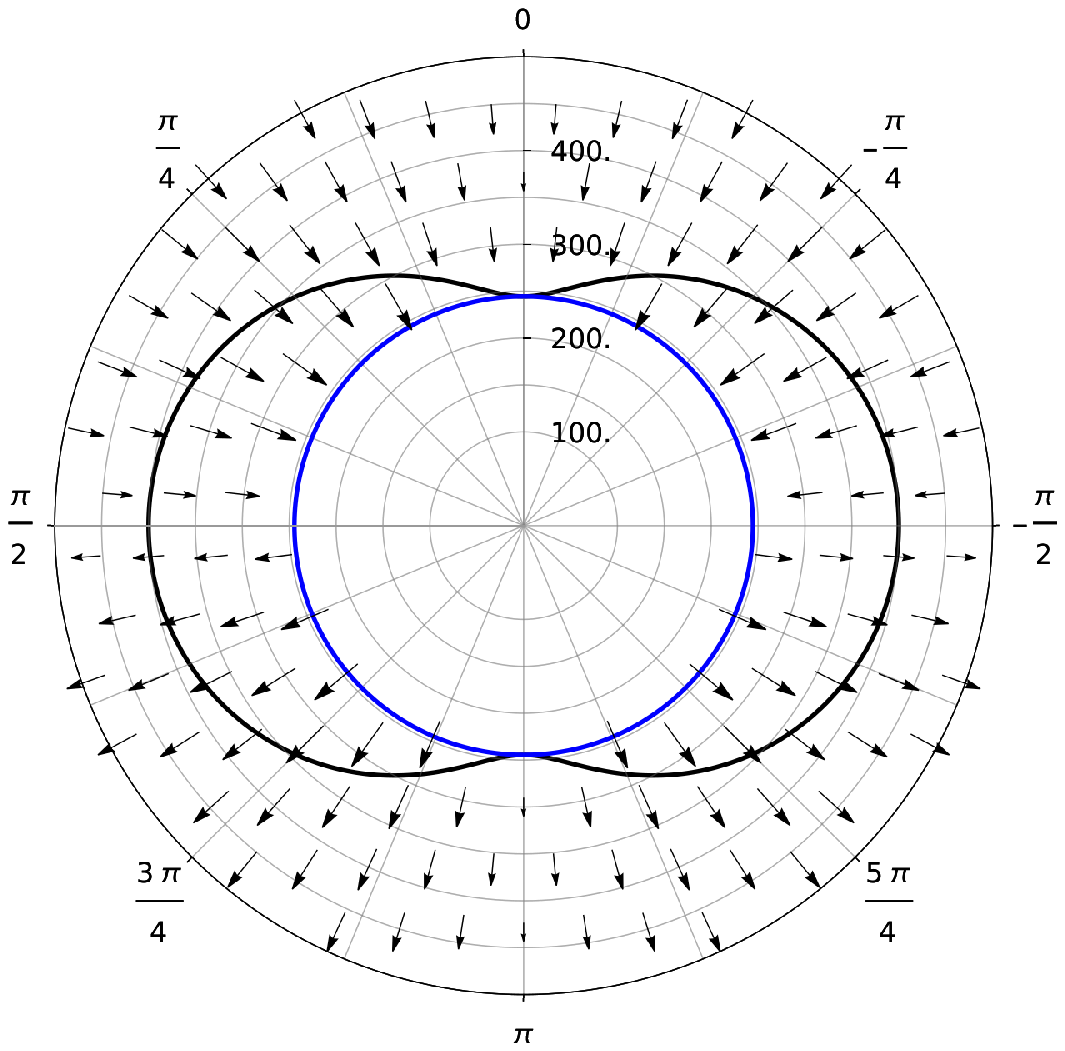}	
			\end{minipage}
			&			
			\includegraphics*[height=5.7cm]{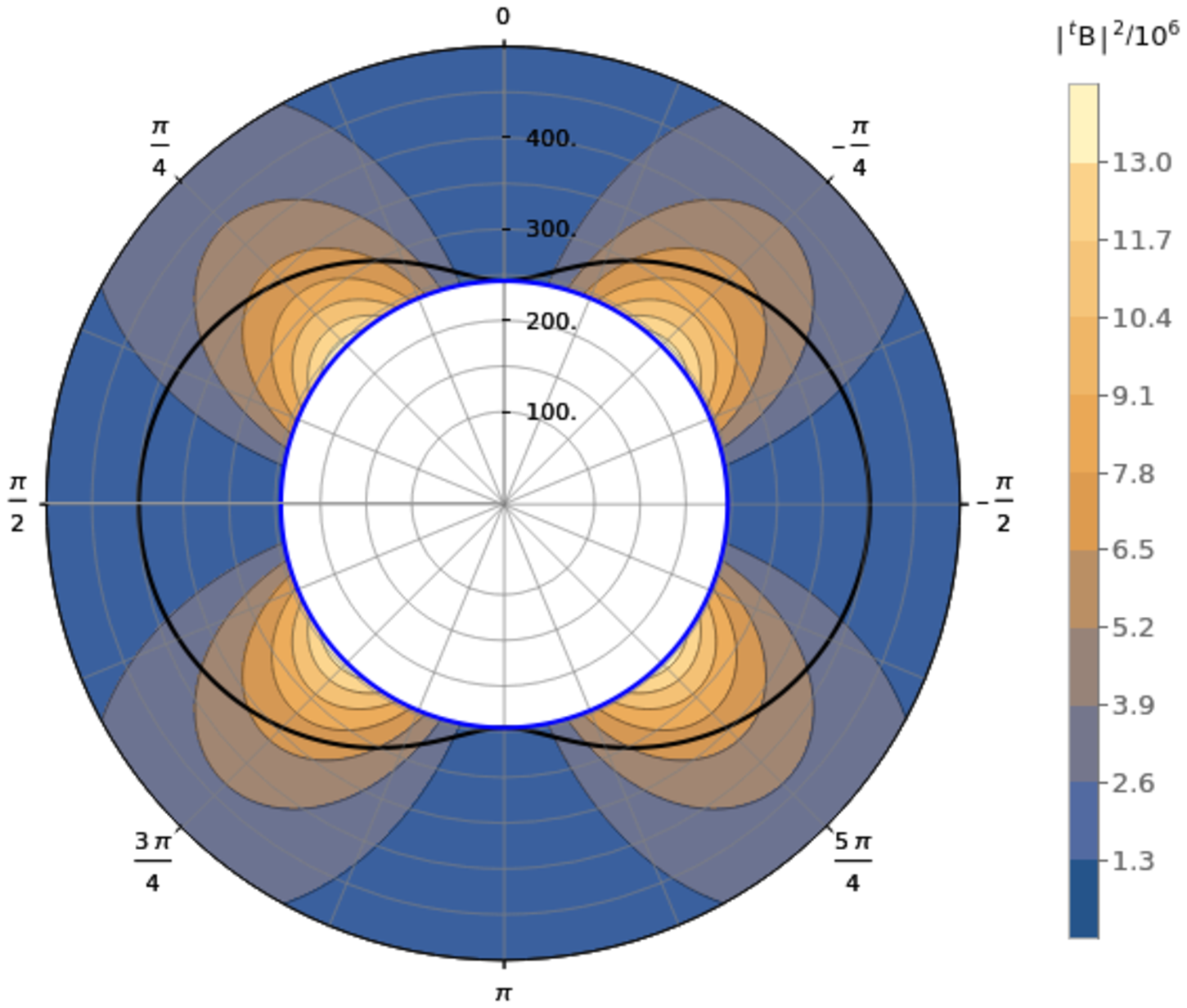}&			
			\includegraphics*[height=5.7cm]{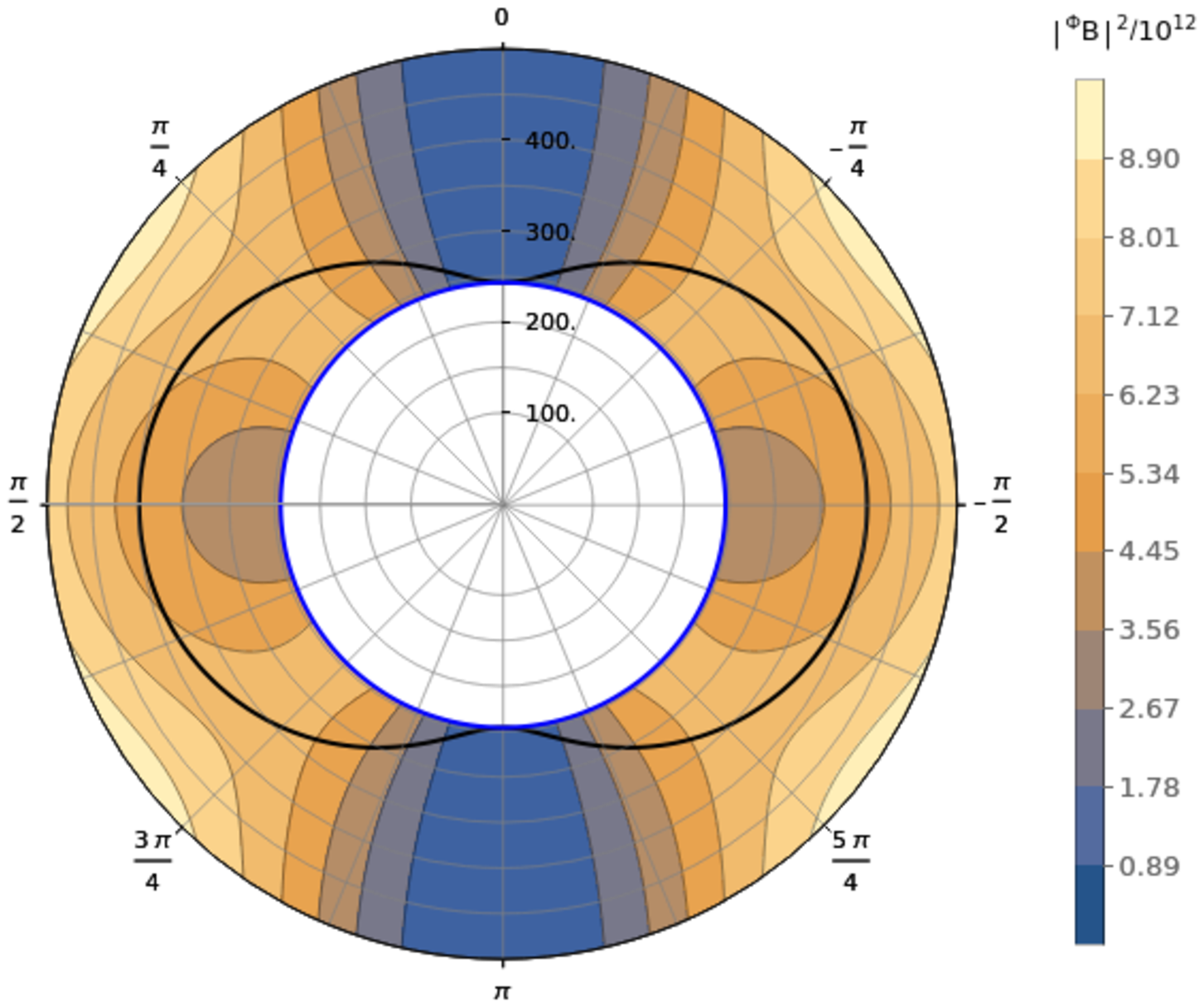} \\ %&
			\begin{minipage}{5cm}
				\vspace{-5.5cm}
			\includegraphics*[height=4.5cm]{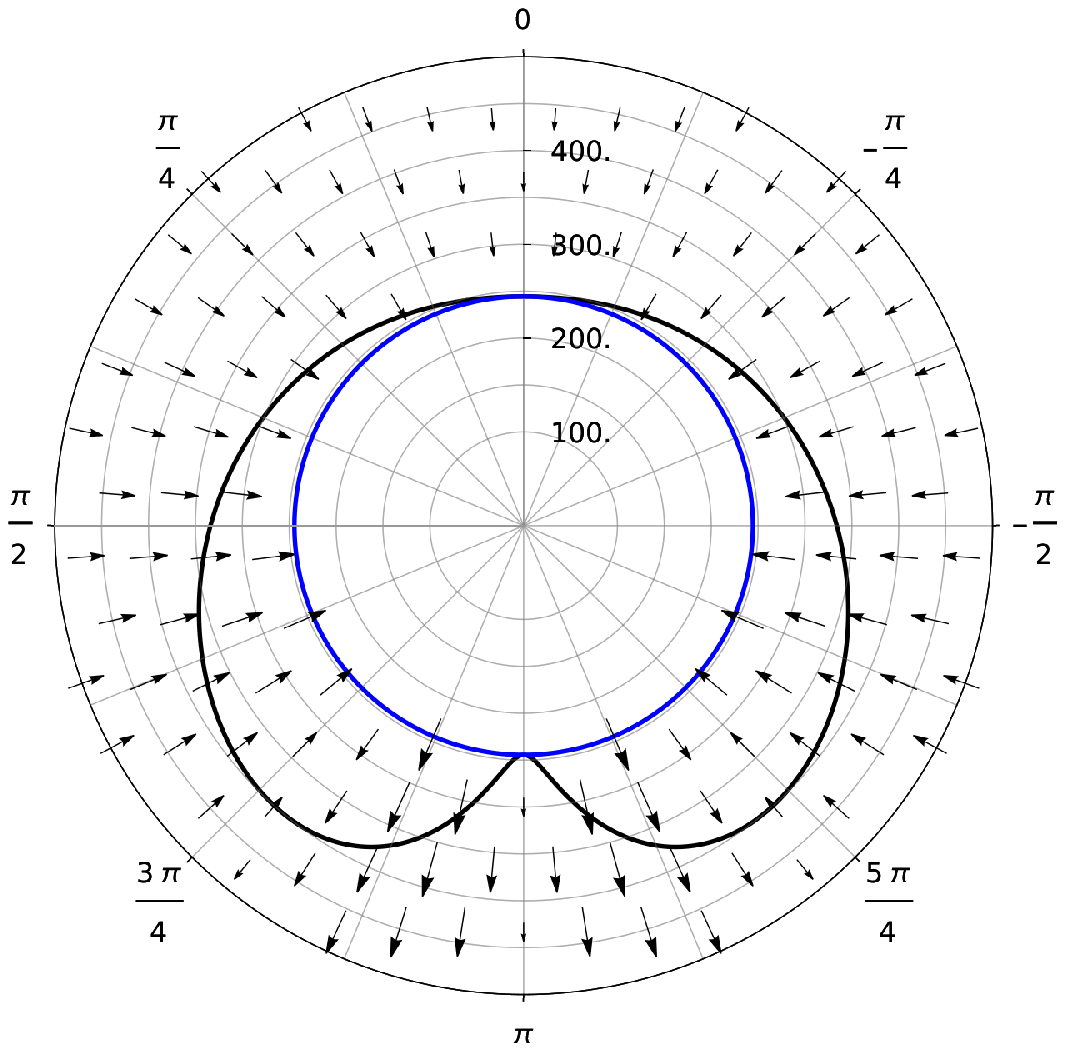}
			\end{minipage}
		    &
			\includegraphics*[height=5.7cm]{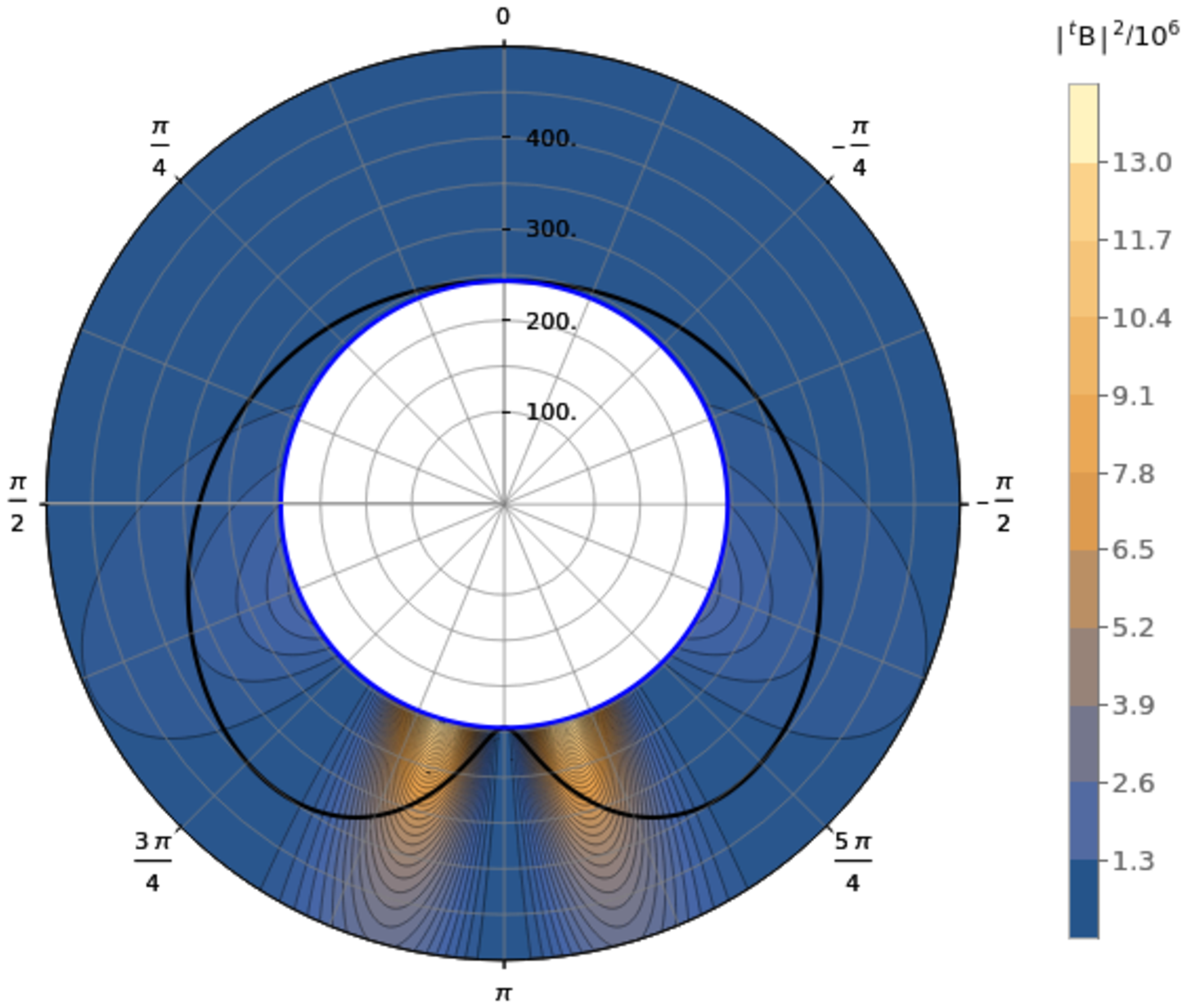}&			
			\includegraphics*[height=5.7cm]{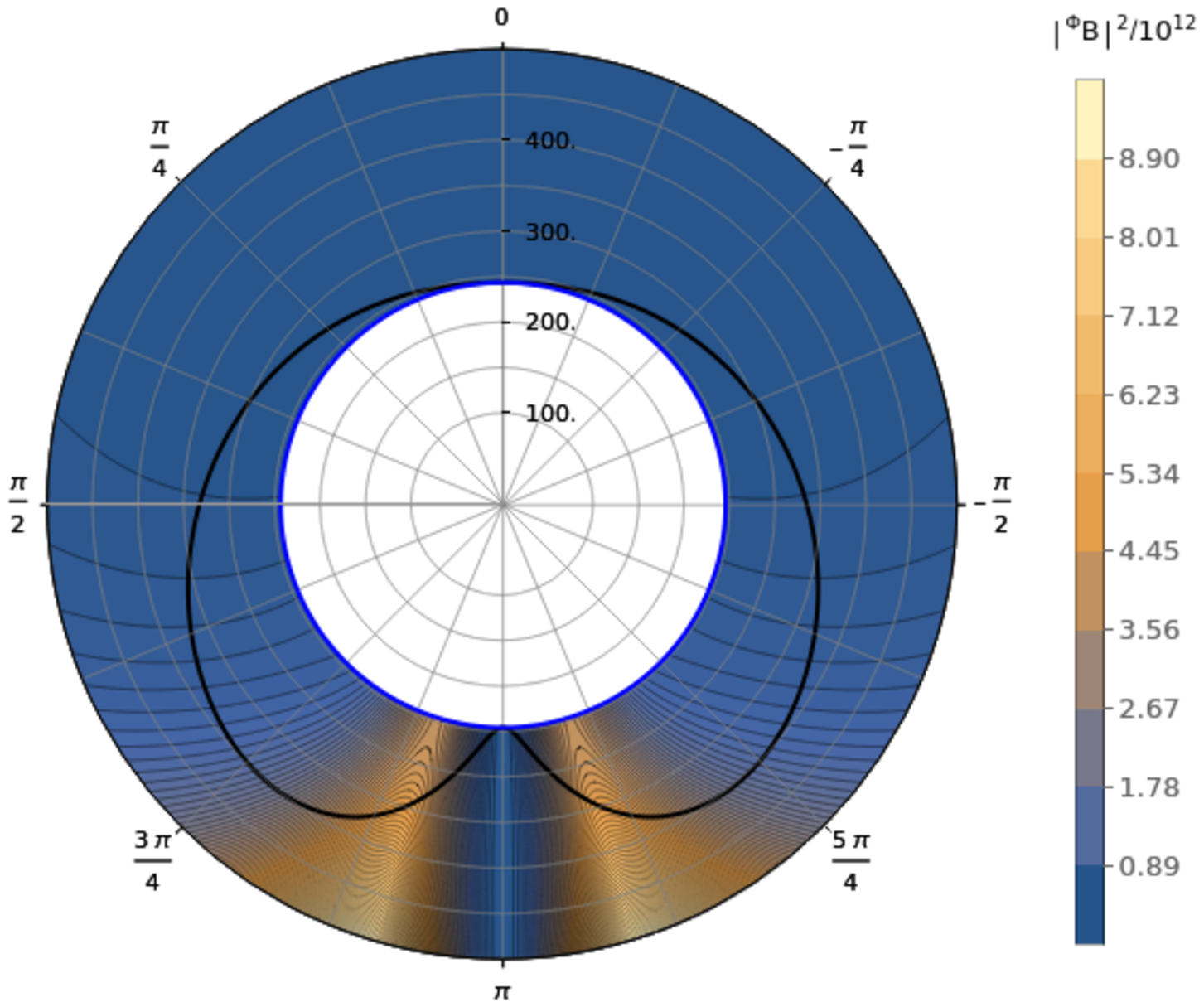}	
		\end{tabular}		
		\caption{Electromagnetic configurations in a boosted Kerr-Schild black hole (\ref{ks})
			as measured by ZAMO observers, assuming a Maxwell test field in the metric background (\ref{ks}).
			The parameters used in these plots are $m_0=200$, $w_0=195$, with boosts respectively $\gamma=0$ (top row)
			and $\gamma=1.0$ (bottom row). A nonzero $\gamma$ deforms the ergosphere as should be expected.
			On the left column we show the magnetic field (\ref{btn}). The small arrows indicate the direction
			of the magnetic lines flowing asymptotically along the $z$-axis in the direction opposite to that of the boost.
			In the middle column we display the modulus of the magnetic fields $|^t{\cal B}|^2$ which is contained purely
			in the plane $(r,\Theta)$. For the boosted case $\gamma=1.0$ (top panel) we observe
			two intense lobes about the direction opposite to that of the boost.
			In the right column we plot the modulus of  $|^\Phi{\cal B}|^2$ for effect of comparison.
			In these plots the color scales characterize the intensity of the modulus
			of the respective magnetic fields in the plane $(r,\Theta)$. We can see that the maximum of the
			flows for both cases $\gamma=0$ and $\gamma=1.0$ are of the same order. However for increasing $\gamma$'s
			the maximum of the modulus of the intensity is more collimated about the $z$-axis,
			in the direction opposite to that of the boost. Although the magnetic flux lines
can be analytically extended beyond 
the event horizon, we did not examined this possibility as the interior is a black hole.
		}
		\label{KS-Btv}
	\end{center}
\end{figure*}
\begin{figure*}
	%\vspace{-20cm}
	\begin{center}
		\begin{tabular}{lll}
			\includegraphics*[height=4.5cm]{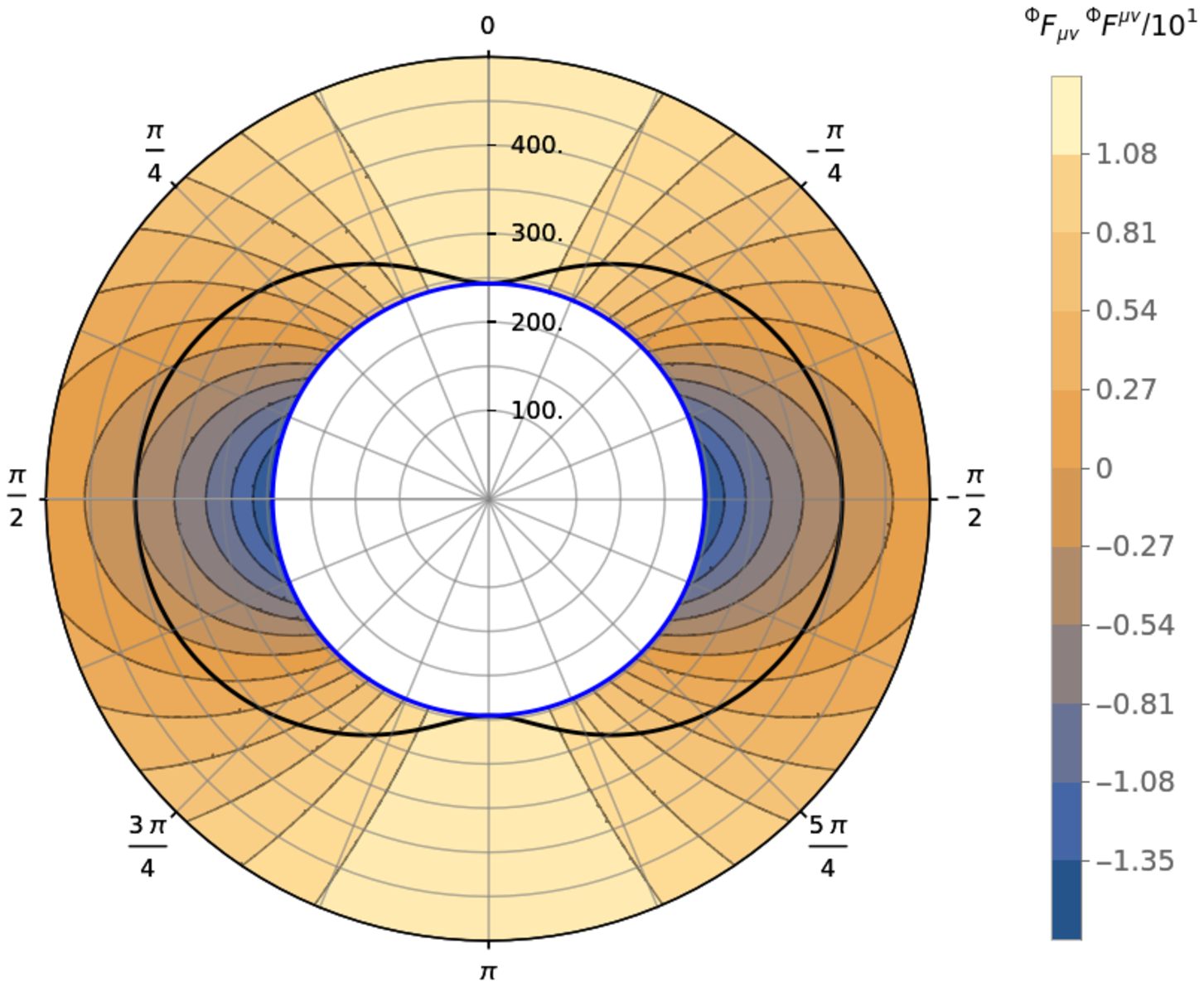}&
			\includegraphics*[height=4.5cm]{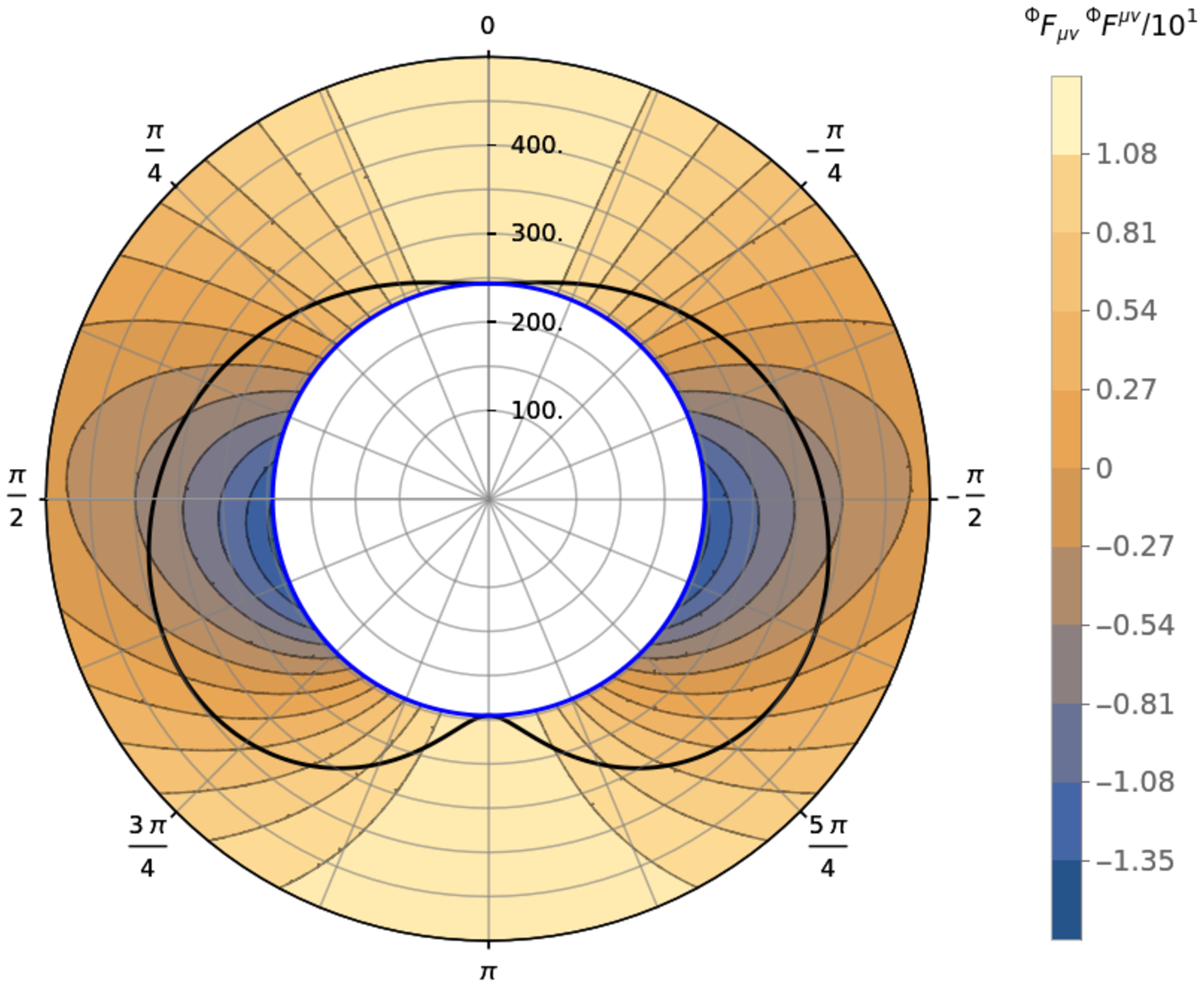}&
			\includegraphics*[height=4.5cm]{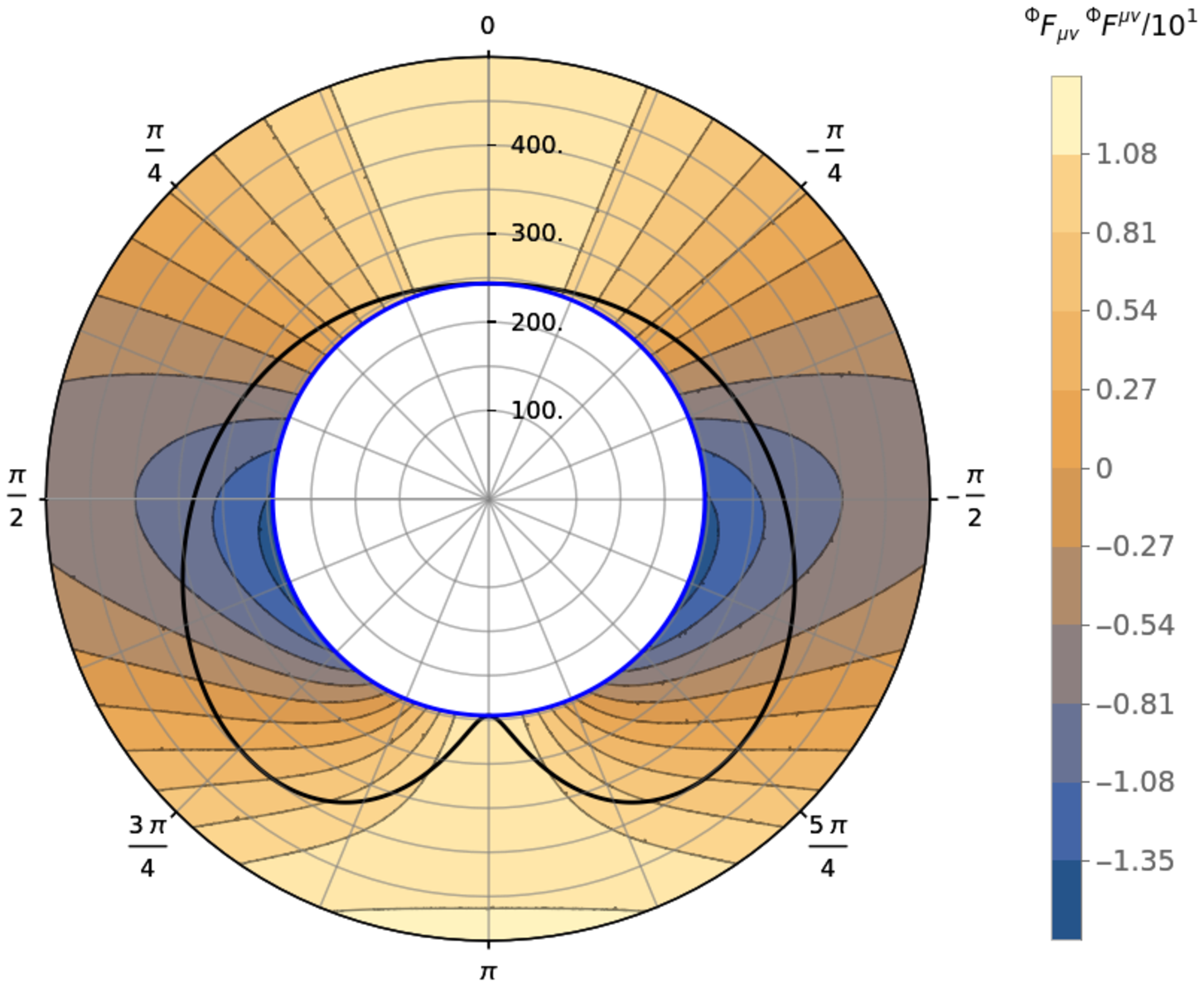}
		\end{tabular}
		\caption{Plots of the intensity of the magnetic field scalar $(^\Phi{ F_{\mu \nu}} ^\Phi {F^{\mu \nu}})$
			in the case of a Kerr-Schild black hole.
			The parameters used for the three plots are $m_0=200$, $w_0=195$, with respective boost parameters $\gamma=0$ (left plot), $\gamma=0.5$ (middle plot)
			and $(\gamma=1.0)$ (right plot). We can see that the magnetic field scalar separates domains of intensity in the plots.
		}
		\label{KS-Kph2}
	\end{center}
\end{figure*}
It is worth remarking that the components of the electric field of the boosted Kerr black holes, given in (\ref{e1}), (\ref{mrt}) and (\ref{ks}), fall with powers of $\mathcal{O}(r^{-2})$ or higher. In this sense, the electric field can be safely neglected.
We must also notice that the coordinate transformations $(u,r,\Theta,\Phi)$ $\rightarrow$ $(t,r,\Theta,\Phi)$ corresponding respectively to
$(\ref{mrt})$ and $(\ref{ks})$ acts as a frame transformation.

\section{The Magnetic Fields in Boosted Rotating Black Holes}
We are now ready to discuss the physical issues of electromagnetic test fields in the gravitational background of a boosted rotating black hole. In the present section we are restricted to the K-S metric (\ref{ks}) with coordinates $(t,r,\Theta,\Phi)$.
As discussed previously these electromagnetic fields are connected to the two Killing vectors of the geometry of the black hole, namely $(\partial/\partial t)$ and $(\partial/\partial \Phi)$, cf. \cite{papapetrou,takahashi}. As mentioned above the associated electric fields can be safely neglected so that the electromagnetic field generated by the geometry (\ref{ks}) is purely magnetic -- the components of such magnetic fields are given by (\ref{mf})-(\ref{mf4}). 

It is worth noticing that the magnetic field connected to the Killing vector $(\partial/\partial t)$ is planar, corresponding to components in the plane $(r, \Theta)$ only, see (\ref{mf})-(\ref{mf1}). Furthermore, remembering that $m_b=\dfrac{m_0}{K^3(\Theta)}$, from (\ref{mf1}) we obtain
\begin{align}
\nonumber
^t{\cal B}^\Phi=&-2\dfrac{K^2}{r}\left(\dfrac{3m_0K'}{K^3}-3\dfrac{m_0K'}{K^4} K\right) =0.
%&0
\end{align}
Hence the only non-null component of the magnetic field $^t\boldsymbol{\cal B}$ is therefore $^t{\cal B}^r$. A further simplification results in
\begin{equation}
\label{btn}
^t{\cal B}^r=-\dfrac{4 m_0 w_0 \sin (\Theta ) (\cosh (\gamma ) \cos (\Theta )+\sinh (\gamma ))}{r (\sinh (\gamma ) \cos (\Theta )+\cosh (\gamma ))^3}.
\end{equation}
%
%%%%%%%%%%%%%%%%%%%%%%%%%%%%%%%%%%%%%%%%%%%%%%%%%%%%%%%%%%%%%%%%%%%%%%%%%%%%%%%%%%%%%%%%%%%%%%%%%%%%%%%
\par In Fig. \ref{KS-Btv} we show the behaviour of electromagnetic configurations in a boosted Kerr-Schild black hole (\ref{ks}) as measured by ZAMO observers.  As seen from above, the electromagnetic field generated by the background geometry is purely magnetic. The top and the bottom rows are distinguished by boost parameters. In fact, the parameters used in these plots are $m_0=200$, $w_0=195$, with boosts respectively $\gamma=0$ (top) and $\gamma=1.0$ (bottom).
On the left panels we show the magnetic field (\ref{btn}). The small arrows indicate the direction of the magnetic lines flowing asymptotically along the $z$-axis in the direction opposite to that of the boost. The arrows plotted in the figures also vary in length and width according to the intensity of the magnetic field at its location. It is to mention that for a boosted configuration (bottom panel, $\gamma=1$) there is a collimation of the magnetic field, marked by bold arrows pointing outward. A change in their direction (inward/outward) is verified when $\Theta$ satisfies
\begin{equation*}
\cos (\Theta )=-\tanh (\gamma ),
\end{equation*}
which happens at approximately $3\pi/4$ and $5\pi/4$.  For $\gamma = 0$ this change occurs in the equatorial plane.
In the middle column of Fig. \ref{KS-Btv} we display the modulus of the magnetic fields $|^t{\cal B}|^2$ contained purely in the plane $(r,\Theta)$. For $\gamma=0$ the square of the magnetic field presents four intense symmetric lobes, the intensity of which is dominant in the ergosphere. For the boosted case $\gamma=1.0$ (bottom panel) we observe two intense lobes about the direction opposite to the boost. This remarkable structure can be responsible for a mechanism by which the BH possibly ejects magnetic energy from the ergosphere. Finally, for effect of comparison we plot the modulus $|^\Phi{\cal B}|^2$ (left column). 
%We remark that the modulus of the amplitude $|^\Phi{\cal B}|^2$ in the boosted case $\gamma=1.0$ is six orders of magnitude higher than the amplitude of $|^t{\cal B}|^2$ in the non-boosted case $\gamma=0$. This highest intensity of the magnetic field is collimated about the opposite direction of the boost (case of a K-S black hole). We finally remark that the magnetic flux lines, due to its character of a test field, extend analytically to the inside of the event horizon. We did not examined this possibility as the interior is a black hole.
We remark that the modulus of the amplitude $|^\Phi{\cal B}|^2$ is six orders of magnitude higher than the amplitude of $|^t{\cal B}|^2$ in both cases: non-boosted $\gamma=0$ (top) and boosted $\gamma= 1$ (bottom). This highest intensity of the magnetic field is collimated about the opposite direction of the boost (case of a K-S black hole). We finally remark that the magnetic flux lines can be analytically extended beyond 
%inside  
%to the 
%inside 
%of 
the event horizon. We did not examined this possibility as the interior is a black hole.

In Fig. \ref{KS-Kph2} we show the plots of the intensity of the magnetic field scalar $(^\Phi{ F_{\mu \nu}} ^\Phi {F^{\mu \nu}})$
for the parameters $m_0=200$, $w_0=195$. The respective boost parameters are $\gamma=0$ (left plot), $\gamma=0.5$ (middle plot)
and $(\gamma=1.0)$ (right plot). We can see that the magnetic field scalar separates domains of intensity in the plots.
For $\gamma=0$ the magnetic field scalar $(^\Phi{ F_{\mu \nu}} ^\Phi {F^{\mu \nu}})$ presents two symmetric lobes. As can be seen from the middle and right panel, apart from the ergosphere deformation such lobes are also bended due to an increasing boost parameter.
%
%
%\newpage
%
\section{The magnetic fields in Bondi-Sachs boosted rotating black holes}

Our previous discussions were based on the Kerr-Schild black hole metric (\ref{ks}) where we examined nonsingular electromagnetic configurations as measured by ZAMOs. There we evaluated the Faraday tensor for both Killing vectors in the Kerr-Schild metric, which turned out to be non-singular and purely magnetic. 
%The electric fields resulted zero (actually, it falls with powers of $\mathcal{O}(r^{-2})$ or higher).
%
As already discussed the Kerr-Schild metric resulted from applying the Bondi-Sachs transformations (\ref{e22}) leading (\ref{e1}) to (\ref{ks}), modulo a transformation from Robinson-Trautman coordinates (\ref{mrt}) to the Kerr-Schild coordinates of (\ref{ks}).
\par
In this section we apply these same transformations in the components of electromagnetic fields. Again, the background geometry generates purely magnetic fiels (electric fields fall with power $\mathcal{O}(R^{-2})$ or higher). In this sense we are in a position to examine the electromagnetic test fields of a Bondi-Sachs boosted rotating black hole (\ref{e1}) described by Bondi-Sachs asymptotic observers at the future null infinity.
In other words, since ZAMOs cannot be defined in the Bondi-Sachs spacetime (\ref{e1}) we use the transformations (\ref{e22}) in evaluating the respective equivalents of the electric and magnetic fields in B-S spacetime. Since the 
Killing vectors $\partial  /\partial U$ and $\partial  /\partial t$ differ only by a $K(\Theta)$ factor, and the azimutal vector fields in both coordinates are the same, we are able to apply (\ref{e22}) to transform the magnetic field from Kerr-Shild to Bondi-Sachs coordinates \cite{mageq}. Thus, the magnetic fields associated to the Killing vectors $\partial/\partial t$ and $\partial/\partial \Phi$ in B-S coordinates result as
\begin{widetext}
\begin{eqnarray}
^t{\cal B}^U&=&\frac{2 \sin (\Theta ) K \left(\omega_b \left(2 m_b \left(\sin (\Theta ) K'+\cos (\Theta )
		K\right)+\sin (\Theta ) K {m_b}'\right)+\sin (\Theta ) K m_b {\omega_b}'\right)}{R},~~~~~~~~~~~~~~~~~~~~~~~~~~~~~~~~~~~~~~\\
^t{\cal B}^R&=&-\frac{2 \sin (\Theta ) \left(\omega_b \left(2 m_b \left(\sin (\Theta ) K'+\cos (\Theta ) K\right)+\sin (\Theta ) K {m_b}'\right)+\sin (\Theta ) K m_b {\omega_b}'\right)}{\text{R} K},
\end{eqnarray}	
and
\begin{eqnarray}
^\Phi{\cal B}^U&=&-11 \sin ^2(\Theta ) K^4 m_b^2 K'+\frac{2 \Sigma (\Theta )^2 \sin ^2(\Theta ) K^2 m_b K'}{R}
-\frac{13 \sin ^2(\Theta ) K^6 m_b^3 K'}{R}+2 R \sin ^2(\Theta) K^2 m_b K'\nonumber\\
&&-\frac{4 \sin ^4(\Theta ) m_b \omega_b^2 K'}{R}-2R^2 \sin ^2(\Theta ) K'-\frac{2 \sin ^4(\Theta ) \omega_b^2 K'}{K^2}
+2 \Sigma (\Theta ) \sin^2(\Theta ) K \Sigma '(\Theta )
+2 \Sigma (\Theta )^2 \sin (\Theta ) \cos (\Theta ) K\nonumber\\
&&-\frac{2 \sin ^4(\Theta ) K \omega_b^2 m_b'}{R} +11 \sin (\Theta ) \cos (\Theta ) K^5 m_b^2
+\frac{2 \Sigma(\Theta ) \sin ^2(\Theta ) K^3 m_b \Sigma '(\Theta )}{R} +\frac{13 \sin (\Theta ) \cos (\Theta ) K^7 m_b^3}{R}\nonumber\\
&&-2 R \sin (\Theta ) \cos (\Theta ) K^3 m_b -\frac{2 \sin ^4(\Theta )K m_b \omega_b \omega_b'}{R}
-\frac{4 \sin ^3(\Theta ) \cos (\Theta ) K m_b \omega_b^2}{R} +2 R^2 \sin (\Theta ) \cos (\Theta ) K
 \nonumber \\
&&+\frac{2 \sin ^4(\Theta )	\omega_b \omega_b'}{K} +\frac{4 \sin ^3(\Theta ) \cos (\Theta ) \omega_b^2}{K},\\
^\Phi{\cal B}^R&=&11 \sin ^2(\Theta ) K^2 m_b^2 K' -\frac{2 \Sigma (\Theta )^2 \sin ^2(\Theta ) m_b K'}{R}
+\frac{13 \sin ^2(\Theta ) K^4 m_b^3 K'}{R} -2 \text{R} \sin ^2(\Theta )m_b K'\nonumber\\
&&+\frac{4 \sin ^4(\Theta ) m_b \omega_b^2 K'}{R K^2} +\frac{2 R^2 \sin ^2(\Theta ) K'}{K^2}+\frac{2 \sin ^4(\Theta ) \omega_b^2 K'}{K^4}
-\frac{2 \Sigma (\Theta ) \sin ^2(\Theta ) \Sigma '(\Theta )}{K}
-\frac{2 \Sigma (\Theta )^2 \sin (\Theta ) \cos (\Theta	)}{K}\nonumber\\
&&+\frac{2 \sin ^4(\Theta ) \omega_b^2 m_b'}{R K} -11 \sin (\Theta ) \cos (\Theta )K^3 m_b^2-\frac{2 \Sigma (\Theta ) \sin ^2(\Theta ) K m_b \Sigma '(\Theta)}{R} -\frac{13 \sin (\Theta ) \cos (\Theta ) K^5 m_b^3}{R} \nonumber\\
&&+2 R \sin (\Theta ) \cos(\Theta ) K m_b + \frac{2 \sin ^4(\Theta ) m_b \omega_b \omega_b'}{R	K} +\frac{4 \sin ^3(\Theta ) \cos (\Theta ) m_b \omega_b^2}{R K} - \frac{2 R^2	\sin (\Theta ) \cos (\Theta )}{K}
-\frac{2 \sin ^4(\Theta ) \omega_b \omega_b'}{K^3}\nonumber\\
&&-\frac{4 \sin^3(\Theta ) \cos (\Theta ) \omega_b^2}{K^3},\\
%\end{eqnarray}
%\begin{eqnarray}
%\\
^\Phi{\cal B}^\Theta&=& -2 \sin ^2(\Theta ) K m_b + \frac{11 \sin ^2(\Theta ) K^3 m_b^2}{R}+\frac{2 R \sin (\Theta )}{K},
%~~~~~~~~~~~~~~~~~~~~~~~~~~~~~~~~~~~~~~~~~~~~~~~~~~~~~~~~~~~~~~~~~~~~~~~~~~~~
\\
^\Phi{\cal B}^\Phi&=&\frac{-5 \sin ^2(\Theta ) m_b \omega_b K' - 2 \sin ^2(\Theta ) K \omega_b 	\omega_b'
-\sin ^2(\Theta ) K m_b \omega_b' - 2 \sin (\Theta ) \cos (\Theta ) K m_b \omega_b}{R}\nonumber\\
&&+\frac{\sin ^2(\Theta ) \omega_b K' -\sin ^2(\Theta ) K
	\omega_b' - 2 \sin (\Theta ) \cos (\Theta ) K \omega_b }{K^2}.
\end{eqnarray}
\end{widetext}
\par
\begin{figure*}
	\begin{center}
		\begin{tabular}{ccc}
			\begin{minipage}{5cm}
				\vspace{-5.5cm}
				{\includegraphics*[height=4.5cm]{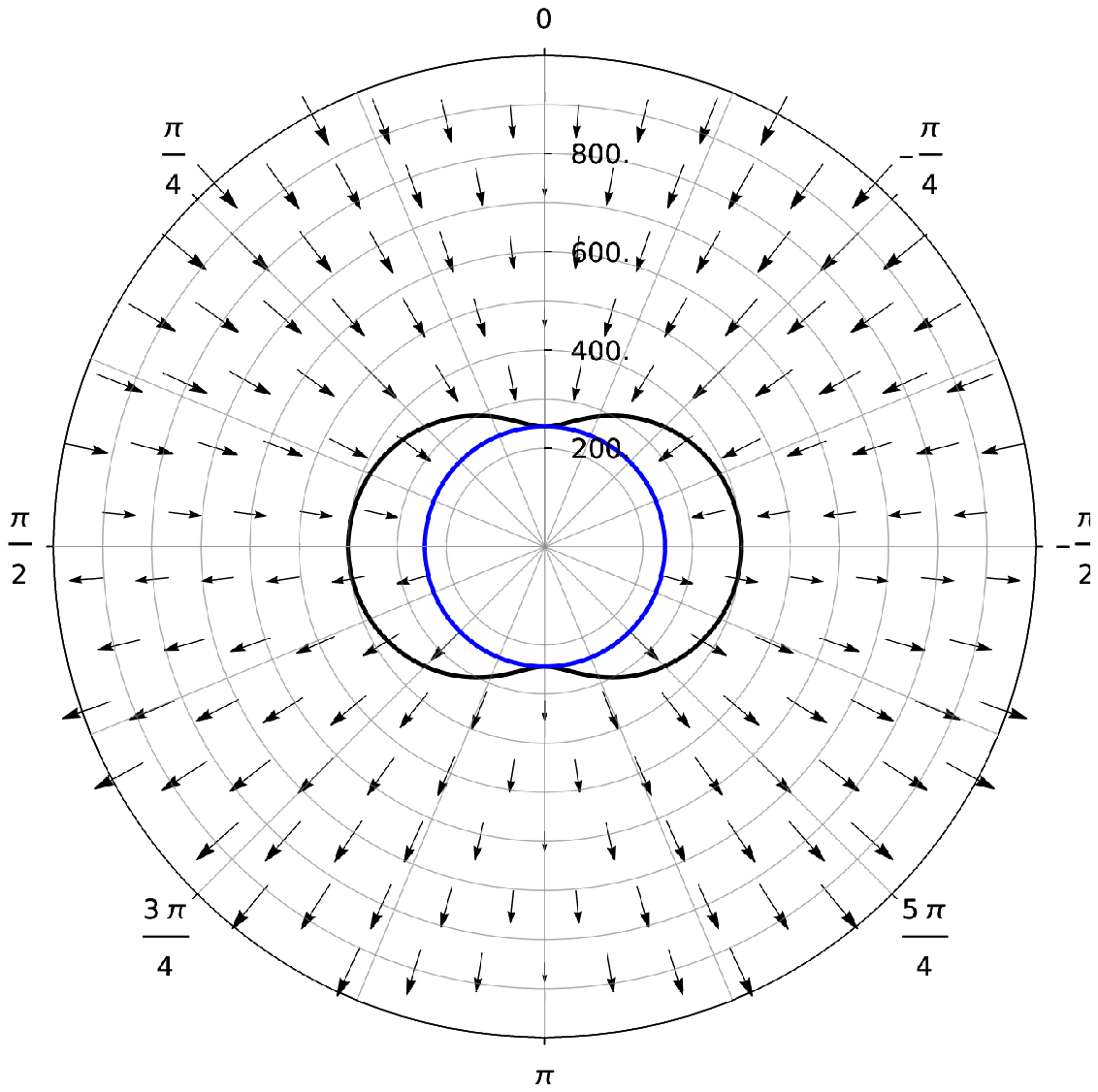}}				
			\end{minipage}&			
			\includegraphics*[height=5.7cm]{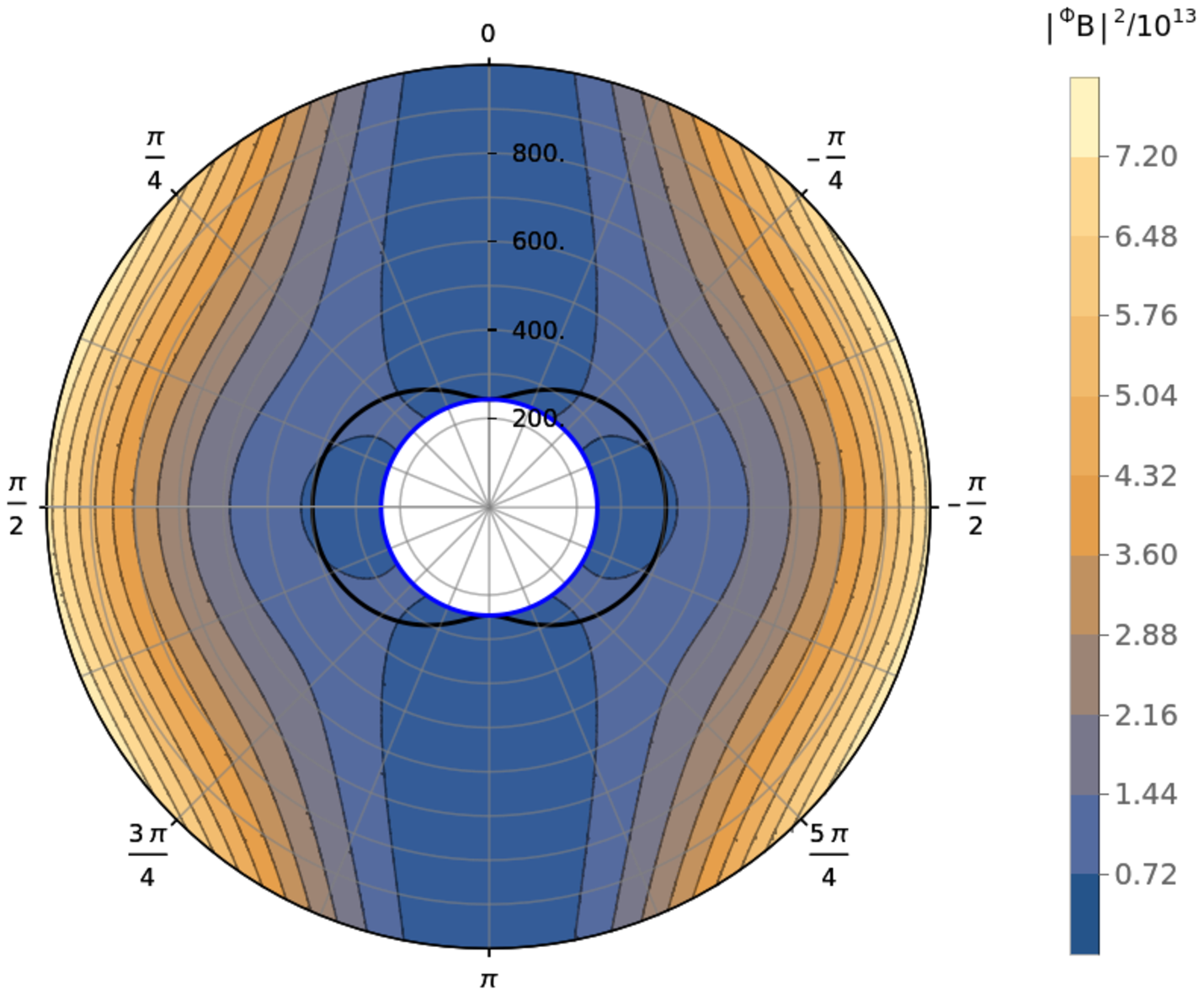}&			
			\includegraphics*[height=5.7cm]{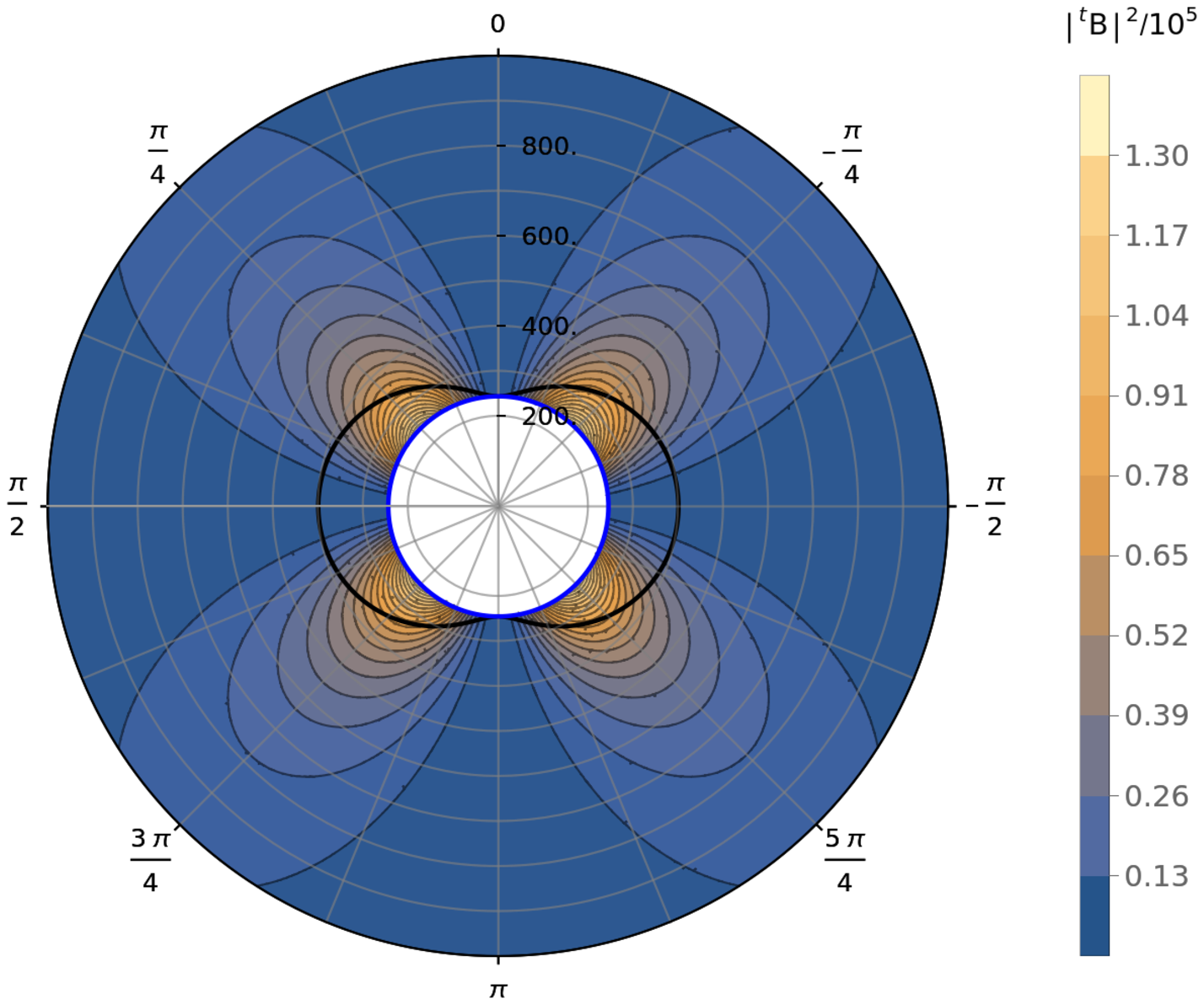} \\ %&
			\begin{minipage}{5cm}
				\vspace{-5.5cm}
			\includegraphics*[height=4.5cm]{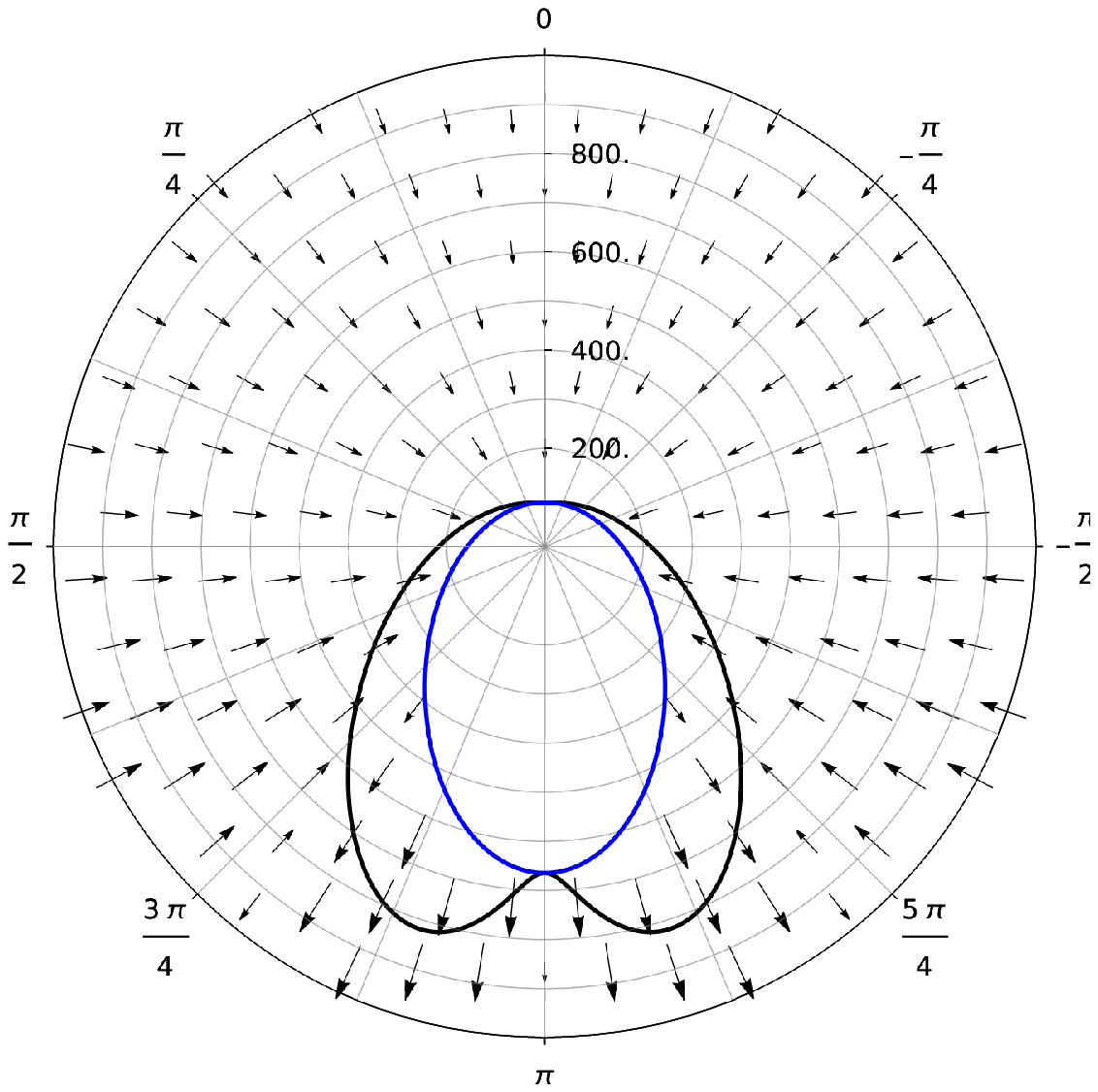} 
			\end{minipage}
			&
			\includegraphics*[height=5.7cm]{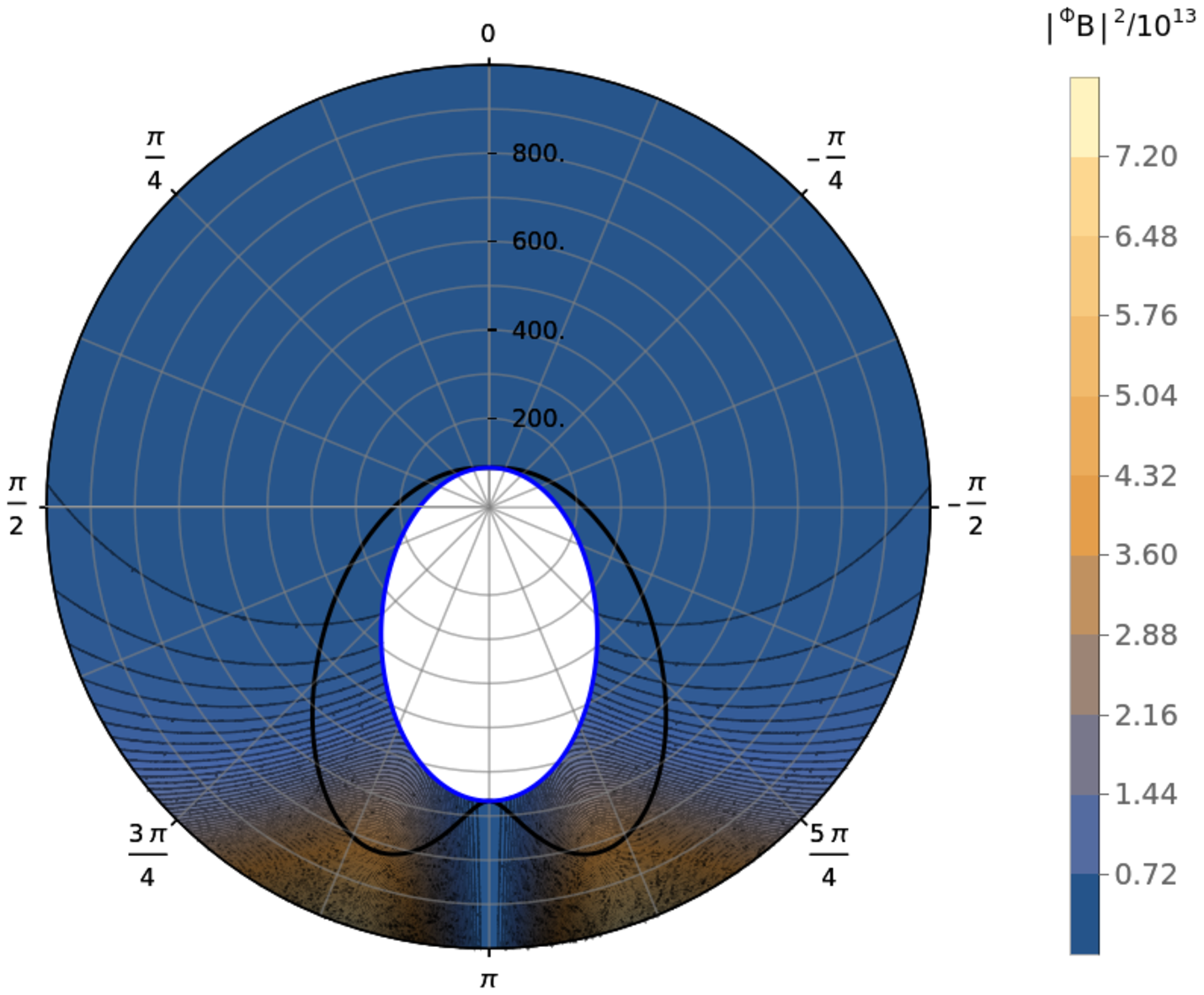} &
			\includegraphics*[height=5.7cm]{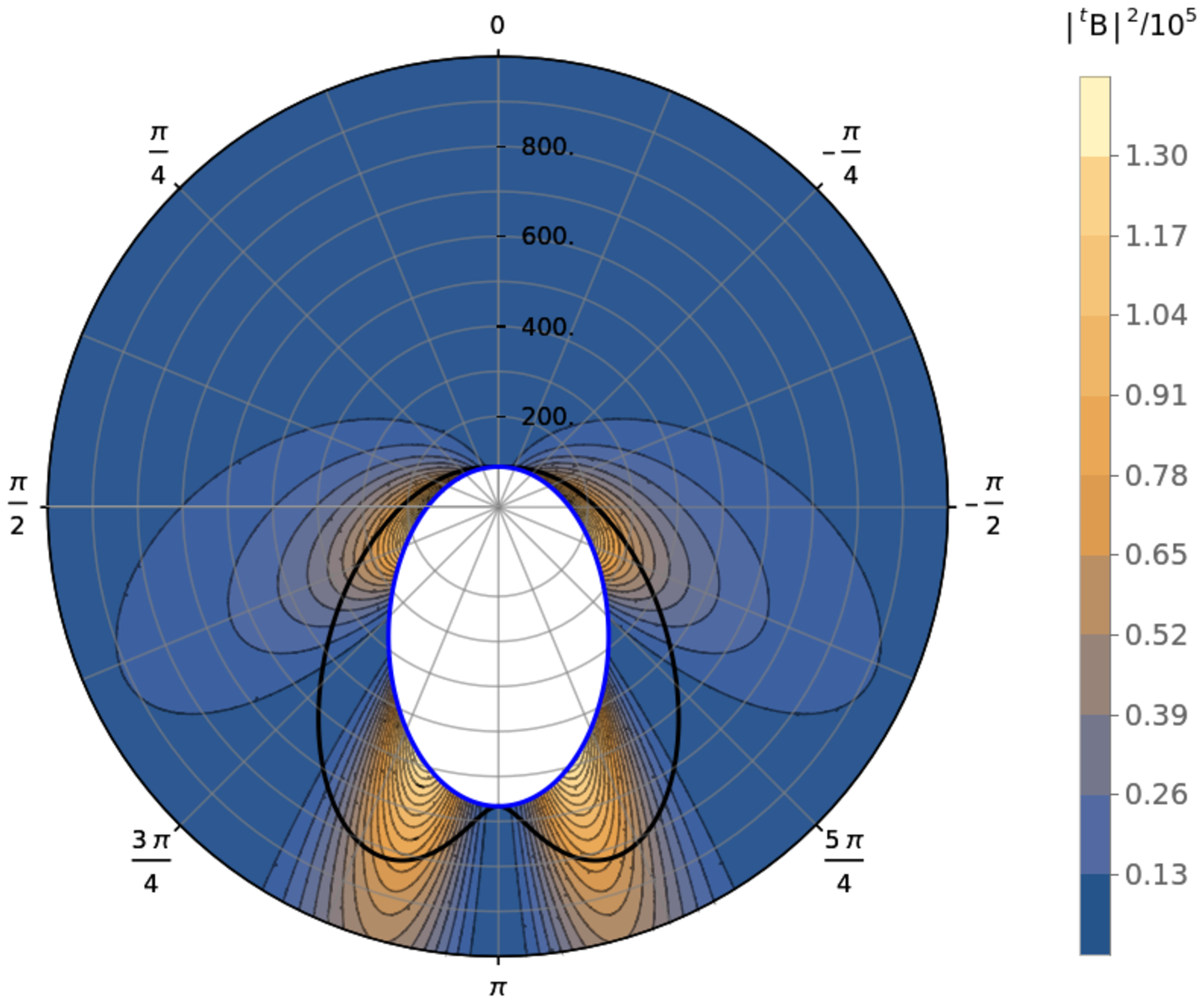}	
		\end{tabular}		
		\caption{Electromagnetic configurations in a boosted Bondi-Sachs black hole (\ref{e1}) with a Maxwell test field. Analogous to the case in (\ref{ks}) the background geometry is purely magnetic. The parameters used in these plots are $m_0=200$, $w_0=195$, with boosts respectively $\gamma=0$ (top row) and $\gamma=1.0$ (bottom row). The qualitative behaviour is analogous to that of the K-S case apart from the fact that a nonzero $\gamma$ deforms the event horizon as well. Again, the magnetic flux lines
can be analytically extended beyond 
the event horizon.        
		}
		\label{BS-Btv}
	\end{center}
\end{figure*}

\begin{figure*}
	\begin{center}
		\begin{tabular}{lll}		
			\includegraphics*[height=5.2cm]{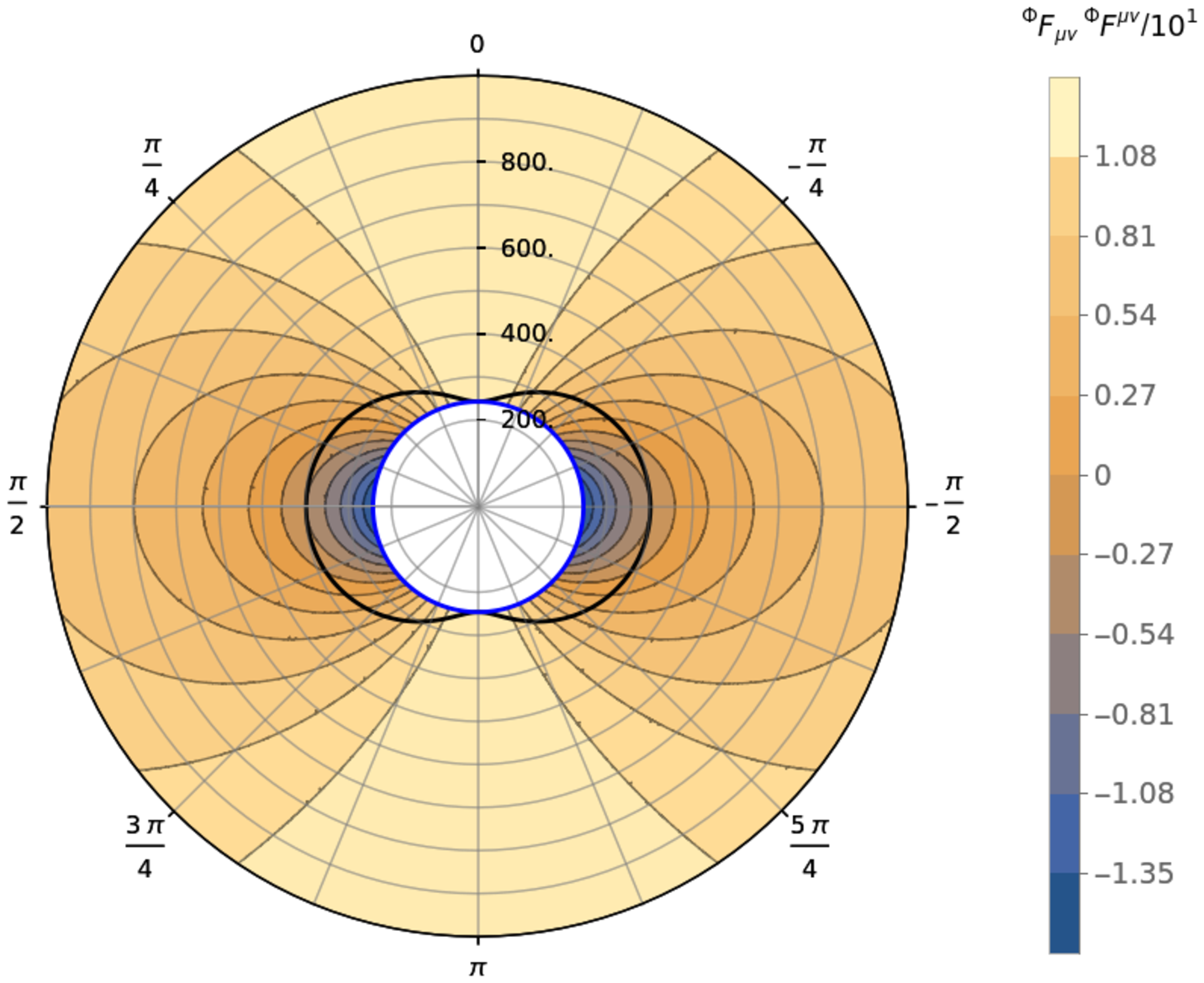}&
			\includegraphics*[height=5.2cm]{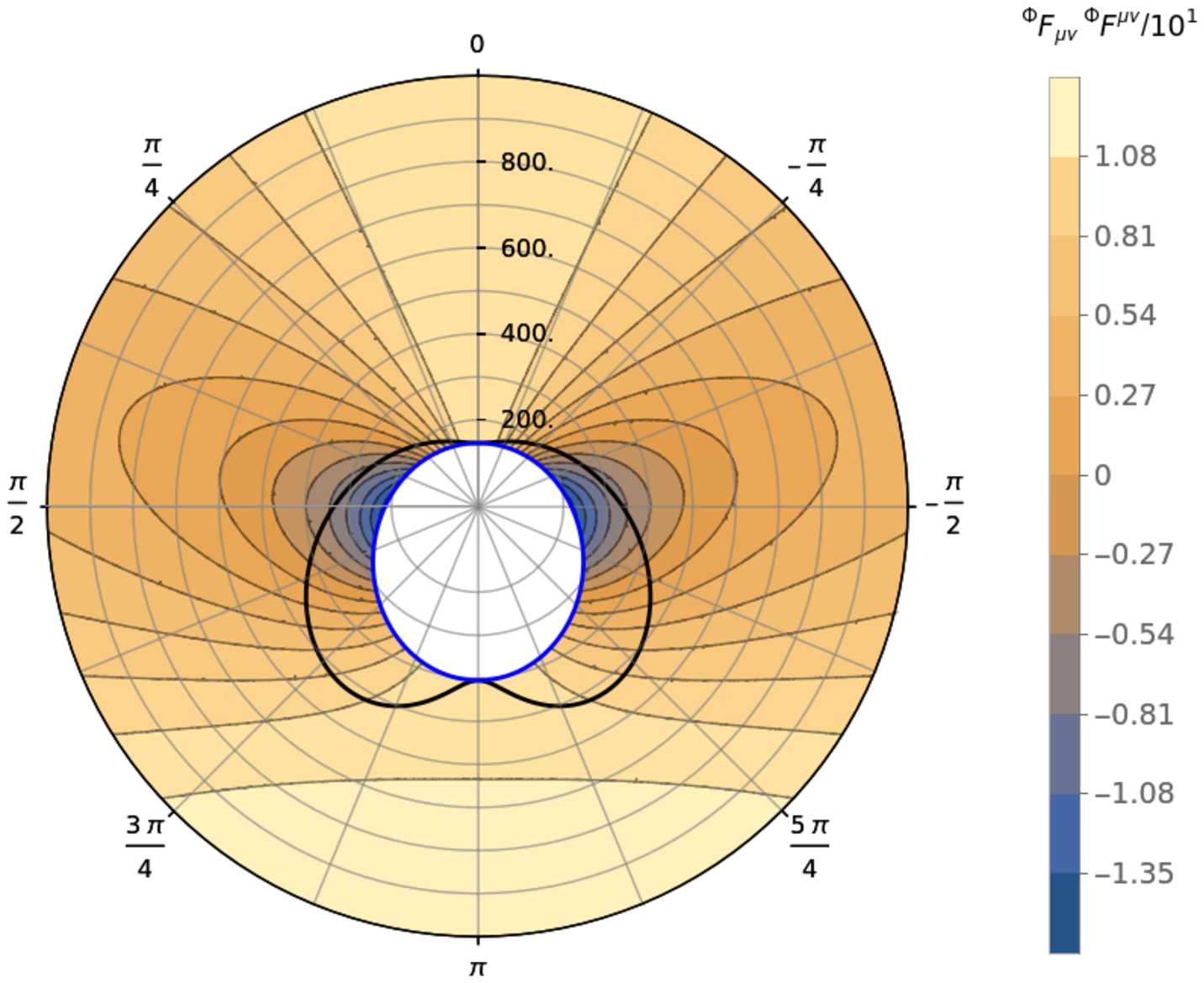}&
			\includegraphics*[height=5.2cm]{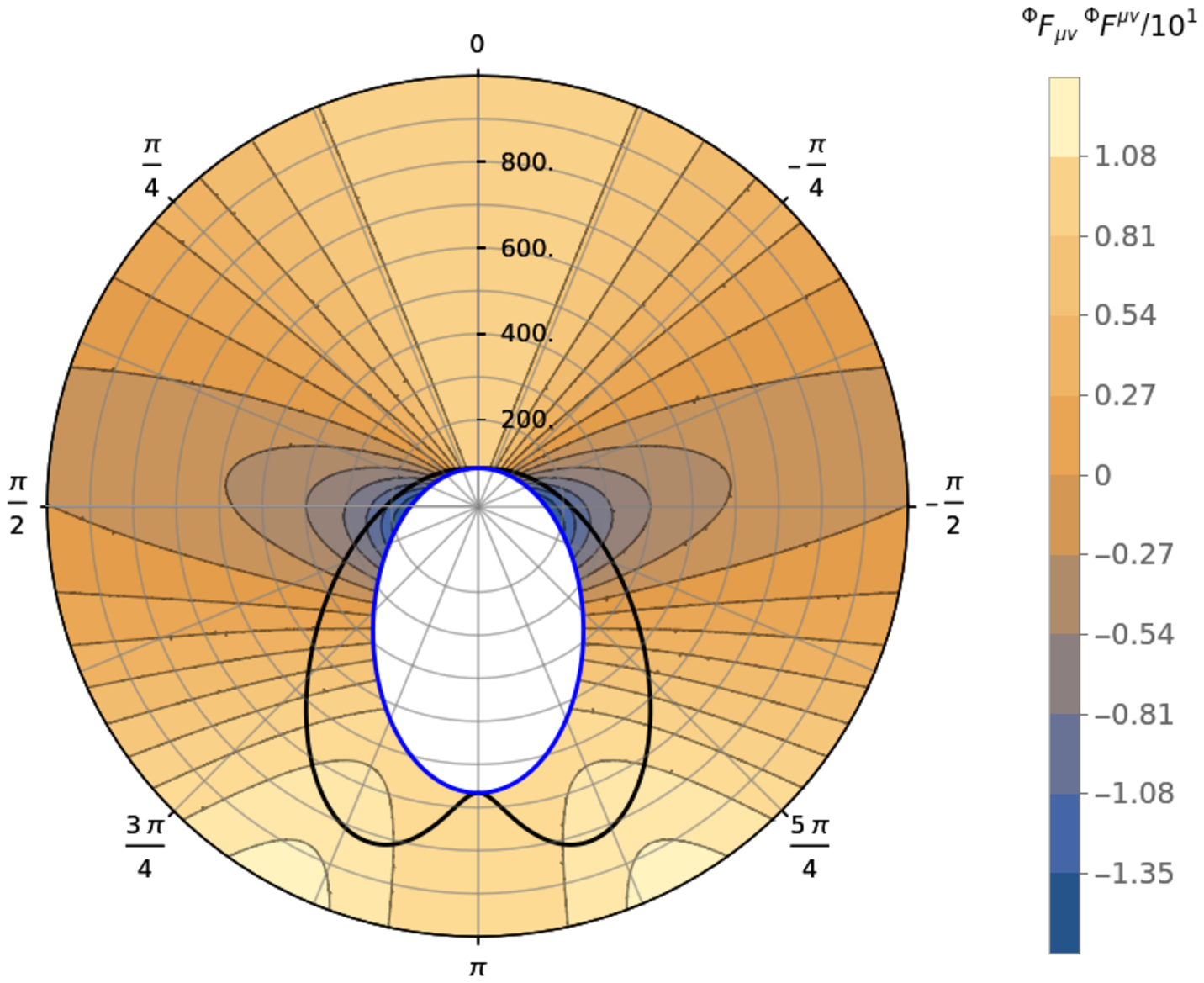}	
		\end{tabular}		
		\caption{Plots of the intensity of the magnetic field scalar $(^\Phi{ F_{\mu \nu}} ^\Phi {F^{\mu \nu}})$ in the case of a Bondi-Sachs black hole. The parameters used for the three plots are $m_0=200$, $w_0=195$, with respective boost parameters $\gamma=0$ (left plot), $\gamma=0.5$ (middle plot) and $(\gamma=1.0)$ (right plot). We can see that the magnetic field scalar separates domains of intensity in the plots.
		}
		\label{BS-Kph2}
	\end{center}
\end{figure*}

 To start, let us consider the Kerr-Schild metric (\ref{ks})
in which the event horizon is defined as $g^{rr}=0$.
%(see Fig. \ref{ergoB2}).
It is a straightforward result that the B-S transformations (\ref{e22}) lead the Kerr-Schild metric (\ref{ks}) to the Bondi-Sachs geometry (\ref{e1}),
where the 2-dim event horizon reads $g^{RR}=0$. In this case the ergosphere and the event horizon are deformed by the presence of the boost as illustrated in Fig. \ref{BS-Btv}. 
In this figure we illustrate electromagnetic configurations in a boosted Bondi-Sachs black hole (\ref{e1}) with a Maxwell test field. Apart from the horizon deformation, here we observe the same qualitative behaviour as in the K-S case. We remark that in our approach we considered the separatrix surface in which the arrows change their direction and obey the same rule as the K-S case. In Fig. \ref{BS-Kph2} we plot the intensity of the magnetic field scalar $(^\Phi{ F_{\mu \nu}} ^\Phi {F^{\mu \nu}})$ in the case of a Bondi-Sachs black hole. The parameters used for the three plots are $m_0=200$, $w_0=195$, with respective boost parameters $\gamma=0$ (left plot), $\gamma=0.5$ (middle plot) and $(\gamma=1.0)$ (right plot). We can see that the magnetic field scalar separates domains of intensity in the plots.
For $\gamma=0$ the magnetic field scalar $(^\Phi{ F_{\mu \nu}} ^\Phi {F^{\mu \nu}})$
presents two symmetric lobes. As can be seen from the middle and right panel, apart from the ergosphere and horizon deformation such lobes are also bended due to an increasing boost parameter.
\par
At this stage we must call attention to the following point. It is well known that General Relativity is a theory of gravitation invariant under diffeomorphisms and this is the reason why quantities like $F_{\mu\nu} F^{\mu\nu}$ are called invariants. However, it is worth noticing that transformations (\ref{e22}) leading (\ref{e1}) to (\ref{ks}) are not diffeomorphisms so that quantities like $F_{\mu\nu} F^{\mu\nu}$ are not invariant under general coordinate transformations.
On the other hand, it can be easily seen from Figs. \ref{KS-Kph2} and \ref{BS-Kph2} that apart from the event horizon deformation, the overall pattern
is slightly the same. Such a feature could be expected once the metric (\ref{ks}) satisfies Einstein field equations in the same scale of energy/curvature as that of (\ref{e1}).
%#############################################################################################################
%
\section{Final Remarks}

\noindent \par In this article we develop a study concerning the electromagnetic fields in the presence of a Kerr-boosted black hole. Here we use a method through the choice of a Kerr-boosted metric\cite{ids1} written in different coordinate systems, namely, Bondi-Sachs, Robinson-Trautman and Kerr-Schild coordinates, in this order. These options gave us a flexible analysis where the several features could be explored. The transformation from B-S into K-S coordinates is necessary to make possible to define a ZAMO. These are the proper observers which measure Maxwell fields around the black hole.
Such electromagnetic fields are engendered from Killing vectors which arise from spacetime isometries. 
It is also important to note that this method considers a metric which comes originally from a general twisting spacetime whose boost (along the symmetry rotation axis) is given by the BMS group\cite{ids1} as it should be in the context of the characteristic initial value problem (CIVP).
\noindent \par The spacetime considered is expanded in a $1/r$ series and, up to the first order, the metric can be shown as a solution of the Einstein equations. Through this spacetime, we constructed the Maxwell fields via isometries up to the order considered and show that all electric field components are null differently of its magnetic counterparts. In this sense we can say we have a case, up to this order, similar to a quasi-neutrality regime as seen in some plasmas. We will explore this in a future work.
\noindent \par For this purely magnetic field written in K-S coordinates, the vector fields are planar and we plot several figures in which some features can be displayed. There is a separatrix which divides the vector field in two regions, each one with opposite directed vectors flowing assymptotically towards the opposite direction of the boost along the $z$-axis. This separatrix changes its direction as one changes the values of the boost parameter from the equatorial plane, where the boost parameter is equal to zero. In a similar manner, we plot the square of the magnetic vector (for both Killing vectors $\partial /\partial t$ and $\partial/\partial \phi$) and we observe that the direction of the lobes depend on the boost parameters, similar to the case mentioned above. The same separatrix discussion can be made again for this second case.
\noindent \par Although it is not possible to obtain the ZAMO in B-S coordinates, it is possible to obtain the components of the magnetic fields through a direct transformation from its K-S components. The electromagnetic scalars (not invariants) are plotted and the magnetic field behaves in a similar manner as seen for the K-S counterparts. The presence of lobes is also noted and its distortion due to the boost parameter is remarkable.
\noindent \par In the near future we plan to investigate more features of our method concerning both electromagnetic aspects as well some astrophysical scenarios. In this sense, we have interest on the description of accelerated particles in this background and analyse the electromagnetic waves emitted in a process via Newman-Penrose formalism in a similar way as seen in\cite{chandra} and some characterization of electromagnetic potentials as in\cite{dolan}. In the astrophysical environment we are also interested in few effects such as the Meissner\cite{meissner} and the Blandford-Znajek\cite{Blandford} processes -- already examined for the Kerr case. It would be also worth to compare our method to some numerical results on plasma simulations of black hole jets\cite{parfrey}.


\begin{thebibliography}{99}
\bibitem{mageq} The transformation of the magnetic field from K-S to B-S coordinates was performed using
$$^t{\overset{(BS)}{B}}\,^{\tilde \mu}=\dfrac{\partial x^{\tilde \mu}}{\partial x^\mu}\,^t{\overset{(KS)}{B}}\,^{\mu}$$
%
where $x^{\tilde{\mu}}=(U,R,\Theta,\Phi)$ and $x^{\mu}=(t,r,\Theta,\Phi)$ are the Bondi-Sachs and the Kerr-Schild coordinates, respectively.
\end{thebibliography}
\end{document}